\documentclass[vecphys]{svmult}

\usepackage{makeidx}         
\usepackage{graphicx}        
\usepackage{psfig}           
\usepackage{multicol}        
\usepackage{url}             
\usepackage[bottom]{footmisc}

\def\xmm{{\sl XMM}-Newton}
\def\chan{{\sl Chandra}}
\def\ed{\dot{E}}
\def\gr{g_{\rm r}}
\def\fp{f_{\rm p}}
\def\ssb{\sigma_{\rm SB}}
\def\tc{\tau_{\rm c}}
\def\lnon{L^{\rm nonth}}
\def\tbb{T^\infty_{\rm bb}}
\def\tbbs{T^\infty_{\rm bb,s}}

\def\rbbs{R^\infty_{\rm bb,s}}
\def\rbbh{R^\infty_{\rm bb,h}}
\def\ts{T_{\rm surf}}
\def\tef{T_{\rm eff}}
\def\tefin{T_{\rm eff}^\infty}
\def\rbb{R^\infty_{\rm bb}}
\def\tpc{T_{\rm pc}}
\def\rpc{R_{\rm pc}}
\def\rin{R^\infty}
\def\tpcc{T_{\rm pc}^{\rm core}}
\def\rpcc{R_{\rm pc}^{\rm core}}
\def\tpcr{T_{\rm pc}^{\rm rim}}
\def\rpcr{R_{\rm pc}^{\rm rim}}
\def\tef{T_{\rm eff}}
\def\lbol{L_{\rm bol}}
\def\lbols{L_{\rm bol,s}}

\def\fbol{F_{\rm bol}}
\def\fbolin{F_{\rm bol}^\infty}
\def\lbolpc{L_{\rm bol}^{\rm pc}}

\def\lnon{L^{\rm nonth}}
\def\Log{{\rm Log}\,}
\newcommand{\gapr}{\raisebox{-.6ex}{\mbox{
$\stackrel{>}{\mbox{\scriptsize$\sim$}}\:$}}}
\newcommand{\lapr}{\raisebox{-.6ex}{\mbox{
$\stackrel{<}{\mbox{\scriptsize$\sim$}}\:$}}}

\makeindex             

\begin{document}

\title*{Thermal emission from isolated neutron stars: theoretical and
observational aspects}

 \titlerunning{Neutron star thermal emission} 

\author{Vyacheslav E. Zavlin} 


\institute{
Space Science Laboratory,
NASA Marshall Space Flight Center, 
SD59, Huntsville, AL 35805, USA; \texttt{vyacheslav.zavlin@msfc.nasa.gov}
}

\maketitle

\begin{abstract}
The possibility for direct investigation of thermal emission from
isolated neutron stars was opened about 25 years ago with
the launch of the first X-ray observatory $Einstein$.
A significant contribution to this study
was provided by $ROSAT$ in 1990's. The outstanding capabilities of the
currently operating observatories, \chan\ and \xmm,
have greatly increased the potential to observe and analyze
thermal radiation from the neutron star surfaces. Confronting observational
data with theoretical models of thermal emission, presumably formed
in neutron star atmospheres, allows one to infer the surface temperatures,
magnetic fields, chemical composition, and neutron star masses and radii.
This information, supplemented with model equations of state and
neutron star cooling models, provides an opportunity to understand
the fundamental properties of the superdense matter in the
neutron star interiors. I review the current status and most
important results obtained from modeling neutron star  thermal emission
and present selectedg \chan\ and
\xmm\ results on thermal radiation from various types of these objects:
ordinary radio pulsars with ages ranging from about 2 kyr to
20 Myr (J1119--6127,
Vela, B1706--44, J0538+2817,
B2334+61, B0656+14, B1055--52, Geminga,
B0950+08, J2043+2740), millisecond pulsars
(J0030+0451, J2124--3358, J1024--0719, J0437--4715), 
putative pulsars (CXOU~j061705.3+222127, RX~J0007.0+7302),
central compact objects in
supernova remnats (in particular,
1E~1207.4--5209), and isolated radio-quiet neutron stars.
\end{abstract}

\section{Brief historical overview} \label{sec:1}


Before the first neutron star was actually discovered as
a radio pulsar\footnote{PSR B1919+21} by 
Jocelyn Bell in 1997 (Hawish et al. 1968), it had been 
predicted that neutron stars, which are thought to
represent the final stage of the stellar evolution, can be 
powerful sources of thermal X-ray emission just because these
elusive objects are
to be hot (Chiu \& Salpeter 1964, Tsuruta 1964), in 
the literal (expected surface temperature $\ts$$\sim$ 1 MK)
and a figurative sense. Remarkably, this idea and the
discovery of the first pulsar became one of motivations for
further developing X-ray astronomy started at the end
of 50's of the twentieth century.
For objective reasons, I am not an expert on the history of
X-ray astronomy,
but I hope nobody would throw a stone at me for saying that
the hunt for thermal emission from neutron stars began
with the launch of the {\it Einstein} observatory
in 1978.
{\it Einstein} detected X-ray emission in
the 0.2--4 keV range from a number of neutron stars
and neutron star candidates (mainly as compact sources
in supernova remnants [SNRs]). Of those, the well-known middle-aged
radio pulsars B0656+14 and B1055-52 and the old pulsar B0950+08,
which emit thermal X-ray radiation,
are discussed in \S\,4. The launch of the $ROSAT$ mission 
sensitive in the 0.1-2.4 keV range  
opened a ``decade of space science'' in 1990's and provided a great
contribution 
in observing X-ray emission from neutron stars.
Speaking of which, the identification of the $\gamma$-ray
source Geminga as a pulsar and, hence, a neutron star (Halpern \& Holt 1992) 
is one of the major results achieved with $ROSAT$.
Those were also supported by observations at energies up to 10 keV
with the $ASCA$ and Beppo$SAX$
X-ray observatories,
with the $EUVE$ satellite covering the 0.07--0.2 keV range,
as well as with $HST$ in optical/$UV$ range. 
Readers interested in 
more details on results from
observations of neutron stars conducted in the last century 
can find them in the historical review by Becker \& Pavlov (2001).
New excellent observational
data on neutron stars collected with two currently operating
powerful X-ray missions, \chan\ and \xmm\ (both launched
in 1999), provide a breakthrough in studying emission properties
of these enigmatic objects. I do not know whether anyone has ever
accurately counted the total number of {\sl isolated}
(i.e., nonaccreting) neutron stars of different types
detected in X-rays, but I believe the number is at least 
sixty\footnote{As at the end of 2006.}. 
Results from a significant fraction of these observations have 
been reviewed by Becker \& Aschenbach (2002),
Pavlov, Zavlin \& Sanwal (2002), Kaspi, Roberts \& Harding (2006), and
Weisskopf et al. (2007).

\begin{figure}
\centerline{\psfig{file=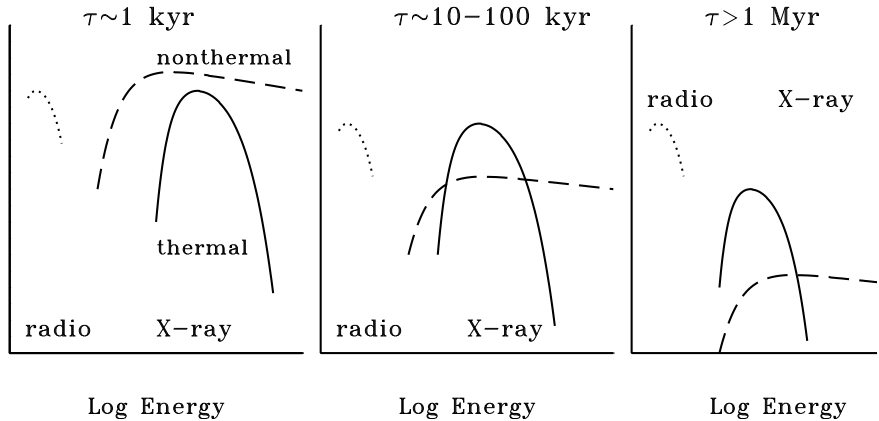,height=6cm,clip=} }
\caption{Sketch illustrating what radiative component,
nonthermal (dashes) or thermal (solid curves), is expected
to dominate in X-ray flux of neutron stars of different
ages $\tau$ (see \S\,2).
}
\label{fig:1}
\end{figure}

\section{Properties of X-ray emission
from isolated neutron stars}

Generally, X-ray radiation from an isolated\footnote{
The term ``isolated'' is omitted hereafter in the text.} 
neutron star
can consist of two distinguished components:
the nonthermal emission due to the pulsar activity
and the radiation originating from the stellar surface.
The nonthermal component is usually
described by a power-law spectral model and attributed to
radiation produced by synchrotron and/or inverse Compton
processes in the pulsar magnetosphere, whereas the thermal
emission can originate from either the entire surface of a cooling
neutron star
or small hot spots around the magnetic poles (polar caps)
on the star surface, or both.
The sketch shown in Figure~1 represents an evolutionary picture
of these two radiative components expected in X-ray emission of
neutron stars. In the majority of very young pulsars ($\tau\sim 1$ kyr)
the nonthermal
component dominates (see the left panel in Fig.~1), making it virtually 
impossible to accurately measure the thermal flux; only upper limits on 
the surface temperature $\ts$ could be derived,
as it was done for the famous Crab pulsar (Tennant et al. 2001)
and PSR~J0205+6449 in the SNR 3C~58 (Slane et al. 2004a).
As a pulsar becomes older, its nonthermal
luminosity decreases
(roughly) proportional to its
spin-down power $\ed=4\pi^2 I P^{-3} \dot{P}$
($I$, $P$, and $\dot{P}$ are the neutron star moment
of inertia, spin period and its derivative, respectively), 
which is thought to
drop with the star age $\tau$, $\dot{E}\sim\tau^{-m}$,
where $m\simeq2$--4 
(depending on the pulsar magneto-dipole braking index).
On the other hand, the thermal luminosity of an aging and cooling
neutron star decreases slower than the
nonthermal one for ages $\tau\sim 10$--100 kyr, up to the end
of the neutrino-cooling era ($\tau\sim 1$ Myr). Thus, the thermal radiation
from the {\sl entire} stellar surface can 
dominate at soft X-ray energies for {\sl middle-aged pulsars}
($\tau\sim 100$ kyr) and some younger pulsars ($\tau\sim 10$ kyr).
This situation is shown in the middle panel of Fig.~1.
For neutron stars older
than about 1 Myr, the surface temperature is
too low, $\ts\lapr 0.1$ MK, to detect
the thermal radiation from the whole surface in X-rays; only hot polar
caps can be observable. As predicted by virtually all pulsar models, these
polar caps can be heated
up to X-ray temperatures ($\sim 1$ MK) by relativistic
particles generated in pulsar acceleration zones.
A conventional assumption about the polar cap radius is that it is close
to the radius within which open magnetic field lines originate from
the pulsar  surface,
$\rpc^*=[2\pi R^3/cP]^{1/2}\simeq 0.5\,[P/0.1\,{\rm s}]^{-1/2}$ km
(for a neutron star radius $R=10$ km).
As the spin-down power $\ed$ is the energy source for both nonthermal
and thermal polar-cap components, it is hard to predict which
of them would prevail in X-ray flux of an old neutron star. However,
it cannot be ruled out (and is proven by observations --- see \S\,4.5) 
that the thermal one may be dominant, as indicated in the right panel
of Fig.~1. 
Remarkably, of neutron stars with detected X-ray emission, 
more than  a half
reveal thermal radiation. 
To interpret these observations, one needs
reliable models of neutron star thermal emission.
This paper reviews theoretical and observational 
aspects of studying thermal radiation from neutron stars.

\section{Theoretical modeling of thermal radiation from neutron
stars} \label{sec:3}

There are a few questions to be answered before immersing into
details on the theoretical modeling of neutron star thermal emission. 

\subsection{Why needed?}

The main question is why studying the thermal emission is needed and 
interesting. Shortly, comparing observed thermal spectrum of a neutron 
star with 
theoretical models can allow one to infer the surface effective
temperature $\tefin$ and total bolometric flux $\fbolin$
(redshifted quantities, i.e.,
as measured by a distant observer) and estimate the actual (unredshifted)
parameters, $\tef=\gr^{-1}\tefin$ and $\fbol=\gr^{-2}\fbolin$,
where $\gr=[1-2 G M/R c^2]^{1/2}$ is the gravitational redshift
determined by the neutron star mass $M$ and radius $R$.
If the distance to the neutron star, $D$, is known, then the
measured temperature and flux yield the 
apparent (redshifted) radius of the star

\begin{equation}
\rin = D\left[\frac{\fbolin}{\ssb(\tefin)^4}\right]^{1/2},
\end{equation}
where $\ssb$ is the Stefan-Boltzmann constant. This in turn links
the actual neutron star radius and  mass 
to each other via the relation $\rin=\gr^{-1} R$, or  

\begin{equation}
M = \frac{c^2 R}{2 G} \left[ 1 - \left(\frac{R}{\rin}\right)^2 
\right]\,\,\,.
\end{equation}

\begin{figure}
\centerline{\psfig{file=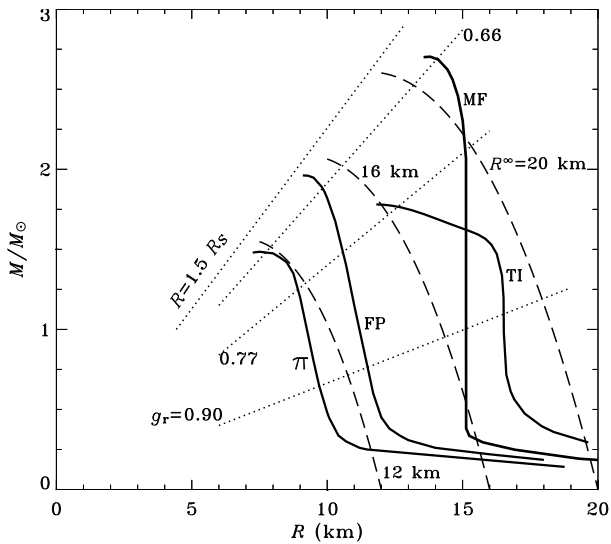,width=10cm,clip=} }
\caption{
Neutron star mass-radius diagram with lines of constant
values of the gravitational parameter $\gr$ (dots), redshifted
radius $\rin=\gr^{-1} R$ (dashes) and four $M(R)$ relations 
(solid curves) corresponding to ``hard'' (MF and TI) and ``soft''
($\pi$ and FP) equations of state of superdense matter
(see, e.g., Shapiro \& Teukolsky 1983).
The values of $M$ and $R$ for realistic equations of state
lie below the straight line $R=1.5 R_{\rm S}$ (or $\gr=1/\sqrt{3}$), where
$R_{\rm S}=2 G M/c^2=2.95 [M/M_\odot]$ km is the Schwarschild radius.
}
\label{fig:2}
\end{figure}

As seen in Figure~2, the latter puts constraints on equation of
state of the superdense neutron star matter.
Moreover,
if one manages to measure the gravitational redshift of a neutron star,
for example, via detecting and identifying spectral features in 
its X-ray flux, then it yields a unique solution for the neutron star
mass and radius, and --- Bingo! --- the long-sought equation of state 
of the neutron star inner matter is found and 
the ultimate goal of the neutron star
physics is achieved! (Un)fortunately, the real life is more complicated
than it may seem. Anyway, even although
solving the neutron star mystery seems
to be far away, investigating thermal emission from these objects
of different ages can trace their thermal evolution, that in turn
sheds light on internal composition and nucleon superfluidity
of the superdense matter (see Yakovlev et al. 2005 for
a review). In addition, inferring surface 
properties of a neutron star (temperature,
magnetic field, chemical composition) tells about
its formation and interaction with environment.

\subsection{More questions}

Like in usual stars, thermal radiation of neutron stars
is formed in the superficial (surface) layers. Hence, the next
question is about the state of the neutron star surface.
In principle, it can be in the gaseous state
(atmosphere) or in a condensed state (liquid or solid),
depending on surface temperature, magnetic field $B$ and chemical
composition.
For instance,
according to the estimates by Lai \& Salpeter (1997),
hydrogen is condensed in surface layers if
$\ts\lapr 0.1$ MK at $B=1\times 10^{13}$ G and
$\ts\lapr 1$ MK at
$B=5\times 10^{14}$ G.
At higher temperatures and/or lower magnetic fields,
hydrogen does not condensate
and forms an atmosphere.
As the majority of known neutron stars seem to possess
surface magnetic fields of $B\sim 10^{10}$--10$^{12}$ G or less,
they are expected to have an atmosphere. Therefore, below I mainly discuss
properties of neutron star atmospheres.

The chemical composition affects not only 
the state of the surface, but it also determines the properties
of emitted radiation. What would the composition of the stellar surface
be? In case of neutron stars, one can expect
that the emitting layers are comprised
of just one, lightest available, chemical element
because heavier elements sink into deeper layers due to
the immense neutron star gravitation (Alcock \& Illarionov 1980).
For instance, even a small amount of hydrogen, with a surface density
of $\sim
1$ g cm$^{-2}$, is sufficient for the
radiation to be indistinguishable from that
emitted from a purely hydrogen atmosphere.
Such an amount of hydrogen, $\sim 10^{-20}M_\odot$, can
be delivered onto the neutron star surface
by, e.g., (weak) accretion from the interstellar medium during
the neutron star life and/or
fallback of a fraction of the envelope ejected
during the supernova explosion.
If no hydrogen is present at
the surface (e.g., because of diffuse nuclear burning ---
see Chang \& Bildsten 2004),
a heavier chemical element
is responsible for
the radiative properties of the neutron star  atmosphere.
However, a mixture of elements can be observed in the emitting layers
if a neutron star is experiencing accretion with such a rate that
the accreting matter is supplied faster than the gravitational
separation occurs.

What else makes neutron star atmospheres very special?
It is of course the enormous gravity at the neutron star surface,
with typical gravitational acceleration $g\sim 10^{14}$-10$^{15}$ cm s$^{-2}$,
and very strong, even huge, surface magnetic 
fields. The gravity makes the atmospheres
very thin, with a typical thickness $H\sim k\ts/[m_{\rm p} g]\sim 0.1$-10 cm
($k$ is the Boltzmann constant and $m_{\rm p}$ is proton mass),
and dense, $\rho\sim 10^{-2}$-10$^2$ g cm$^{-3}$.
Such a high
density causes
strong nonideality effects (pressure ionization,
smoothed spectral features) which must be taken into account
(e.g., Pavlov et al. 1995).
In addition, the strong gravitational field
bends the photon trajectories near the neutron star surface
(Pechenick, Ftaclas \& Cohen 1983),
as illustrated in Figure~3.
This effect depends on the gravitational parameter
$\gr$, and it can even
make the whole surface visible if the neutron star is massive enough,
$1.92\,[10\,{\rm km}/R]<[M/M_\odot]<2.25\,[10\,{\rm km}/R]$. In particular,
the gravitational bending
strongly affects the observed pulsations
of thermal emission
(Zavlin,  Shibanov \& Pavlov 1995). As shown by Pavlov \& Zavlin (1997) and
Zavlin \& Pavlov (1998, 2004a), analyzing pulsed thermal
radiation can put constraints on the mass-to-radius
ratio, $M/R$, and the neutron star geometry (orientation
of spin and magnetic axes with respect to each other and 
direction to a distant observer --- see Fig.~3).

\begin{figure}
\centerline{\psfig{file=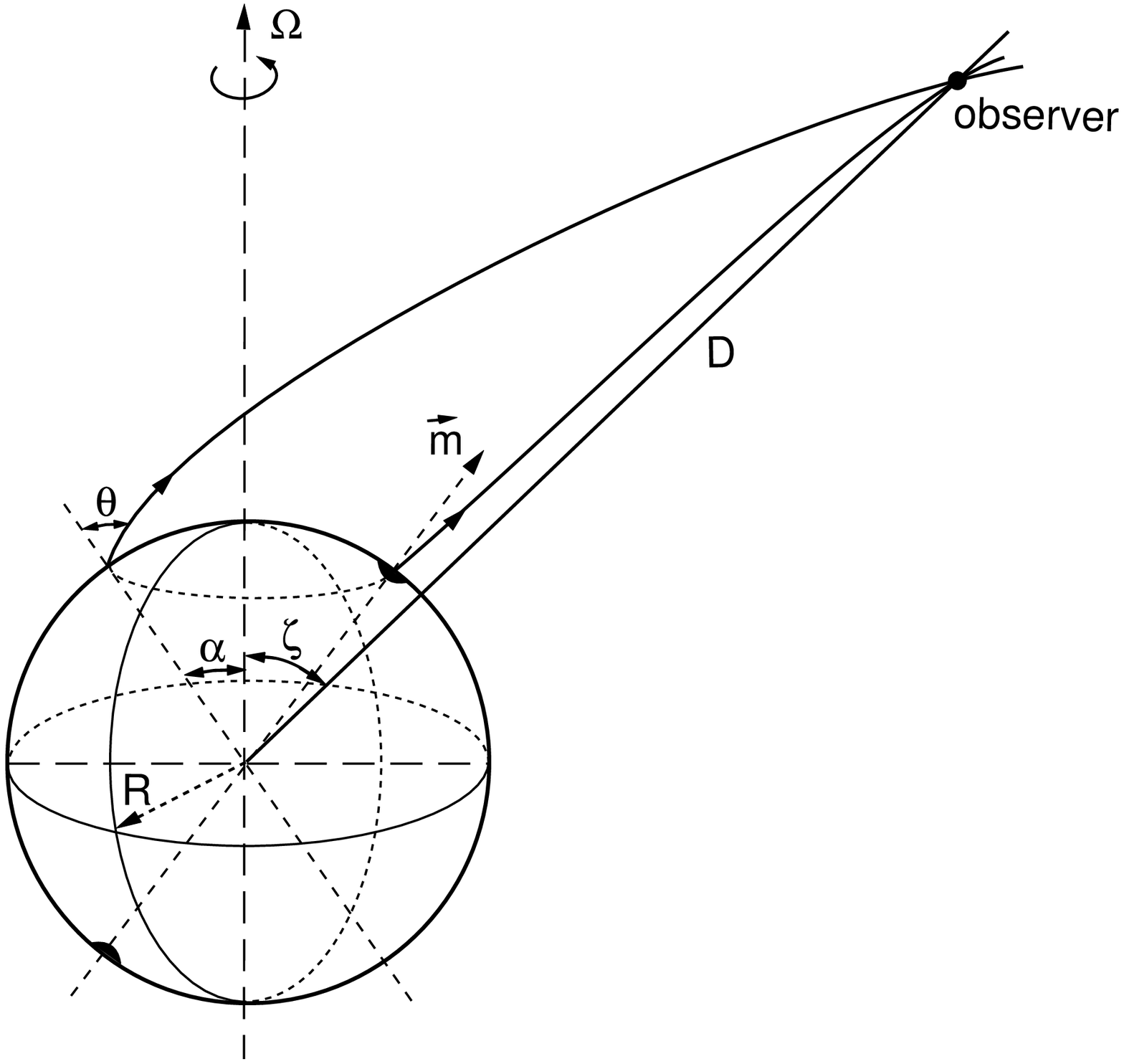,width=8cm,clip=} }
\caption{
Sketch illustrating bending of photon trajectories
near the surface of a neutron star with spin, $\vec{\Omega}$,
and magnetic, $\vec{m}$, axes.
Black spots around the magnetic poles indicate possible
heated polar caps on the star surface.
}
\label{fig:3}
\end{figure}

Huge magnetic fields, up to $B\sim 10^{14}$ G, expected in the surface
layers of neutron stars change
the properties of the atmospheric
matter and the emergent radiation very drastically.
Strongly magnetized atmospheres are essentially anisotropic,
with radiative opacities depending on the magnetic field and
the direction and polarization
of radiation. Moreover, since the ratio of the cyclotron
energy, $E_{\rm ce}=\hbar eB/m_{\rm e}c$, to the Coulomb energy
is very large (e.g.,
$\beta\equiv E_{\rm ce}/[1\, {\rm Ry}] = 850\, [B/10^{12}\, {\rm G}]$
for a hydrogen atom),
the structure of atoms is strongly distorted
by the magnetic field. For instance, the binding (ionization) energies
of atoms are increased by a factor of $\sim \ln^2\beta$ (e.g.,
the ionization potential of a hydrogen atom is
about
0.3 keV at $B=10^{13}$ G).
This in turn significantly modifies ionization
equilibrium of the neutron star atmospheric plasma.
Another important effect is that
the heat conductivity of the neutron star crust is anisotropic, being higher
along the magnetic field. This results in a nonuniform surface
temperature distribution (Greenstein \& Hartke 1983), which
leads to pulsations of the thermal radiation
due to neutron star rotation.

Depending on the magnetic field strength, models of neutron star
atmospheres are differentiated
 in two groups, ``nonmagnetic'' and ``strongly
magnetized''.
The nonmagnetic models are constructed for $B\lapr 10^9$ G,
when the electron
cyclotron energy,
$E_{\rm ce}\lapr 0.01$ keV, is lower than the binding energy
of atoms and thermal energy of particles, $E\sim k\ts$.
As a result, the effect
of the magnetic field on the radiative opacities
and emitted spectra is negligible
at X-ray energies, $E\gapr 0.1$ keV.
These models, constructed assuming $B=0$ G,  
can be applicable to, for example, millisecond pulsars 
and neutron star transients in quiescence
(e.g., Rutledge et al. 1999, 2001a,b and  2002),
whereas magnetized models are intended mostly for radio pulsar with
$B\sim 10^{10}$--10$^{14}$ G. Below I summarize main results 
obtained from these two groups of neutron star atmosphere
models. More details can be found in the extended review by
Zavlin \& Pavlov (2002).

\subsection{Nonmagnetic atmosphere models}

\begin{figure}
\centerline{\psfig{file=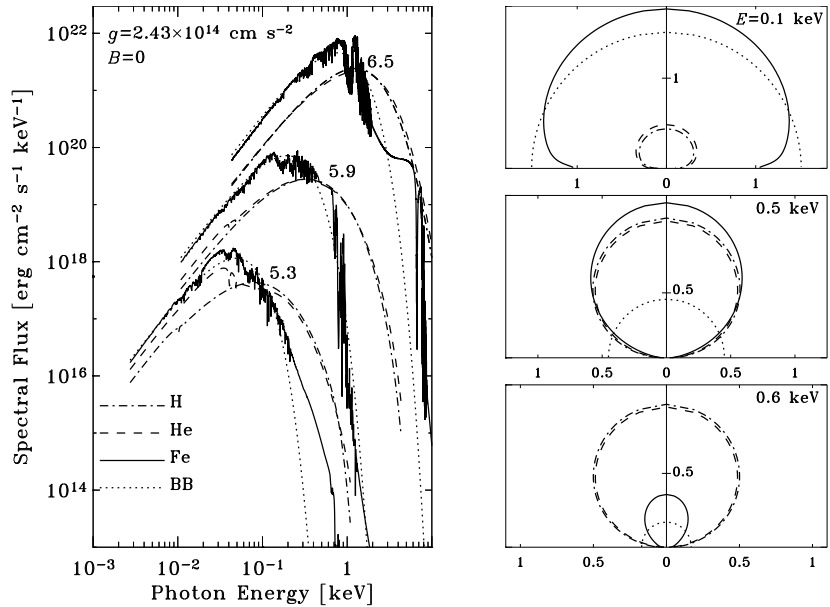,height=8cm,clip=} }
\caption{
{\sl Left}: Spectra of emergent radiation for pure
hydrogen, helium, and iron nonmagnetic atmospheres
with different effective temperatures (numbers near
the curves label $\Log\tef$ [$\tef$ in K]).
`BB' stands for blackbody spectrum.
{\sl Right}: Polar diagrams of normalized spectral specific intensities
at different photon energies, $E$, and 
$\Log\tef=5.9$, and for the same chemical compositions.
The normal to the surface is directed upward.
}
\label{fig:4}
\end{figure}

Modeling nonmagnetic neutron star atmospheres was started in
the pioneering work by Romani (1987). 
Since then,
models for various surface chemical compositions have been
developed by Rajagopal \& Romani (1996), Zavlin, Pavlov \& Shibanov (1996),
Zavlin et al. (1996),
Werner \& Deetjen (2000), Pavlov \& Zavlin (2000a),
Pons et al.\ (2002), G\"ansicke, Braje \& Romani (2002), and
Heinke et al. (2006).

The general approach of the atmosphere modeling is as follows.
Very small 
thickness of a neutron star atmosphere, $H\ll R\approx 10$ km (\S\,3.2), 
allows one to use the plane-parallel (one-dimensional)
approximation.
In addition, because of rather high
densities of the surface layers, the
atmospheres are expected to be in the local thermodynamic equilibrium.
The atmosphere modeling involves
solving three main
equations. The first one is the radiative transfer equation
for the specific spectral intensity $I_\nu$ (e.g., Mihalas 1978):
\begin{equation}
\mu\frac{{\rm d}}{{\rm d}y} I_\nu = k_\nu (I_\nu - S_\nu)\, ,
\end{equation}
where 
$\nu$ is photon frequency,
$\mu$ is cosine of the angle $\theta$ between the normal to
the surface and the wave-vector of outgoing radiation,
$y$ is the column density (${\rm d}y=\rho\, {\rm d}z$, with
$z$ being the geometrical depth),
$k_\nu=\alpha_\nu+\sigma_\nu$ is
the total radiative opacity which includes
the absorption,
$\alpha_\nu$, and scattering, $\sigma_\nu$, opacities,
$S_\nu=(\sigma_\nu J_\nu+\alpha_\nu B_\nu)k_\nu^{-1}$ is the source
function,
$J_\nu=\frac{1}{2}\int_{-1}^1 I_\nu {\rm d}\mu$ is the mean spectral
intensity, and $B_\nu$ is the Planck function.
The boundary condition for this equation is
$I_\nu=0$ for $\mu<0$ at $y=0$,
assuming no incident radiation
at the surface (valid at $R>1.5 R_{\rm S}$ --- see Fig.~2).

The atmospheres are supposed to be in radiative
and hydrostatic equilibrium. The first condition implies that
the total energy flux through the atmosphere is constant
and transferred solely by radiation (electron heat conduction and convection
are of no imortance for typical parameters of interest),
\begin{equation}
\int_0^\infty {\rm d}\nu\int_{-1}^1\, \mu I_\nu\, {\rm d}\mu
=\sigma_{\rm SB}\, \tef^4\, .
\end{equation}
The second condition means that the atmospheric pressure is $p=g\, y$
(the radiative force is insignificant unless $\tef\gapr 10$ MK).
Finally, these three equations are supplemented with the equation
of state for the atmospheric plasma and equations of ionization
equilibrium. The latter is needed for computing the electron number density
and the fractions of ions in different stages of ionization
to obtain the radiative opacity with account
for free-free, bound-free and bound-bound
transitions.

The main results of the atmosphere modeling are
the properties of the emergent radiation demonstrated in Figure~4.
The left panel of Fig.~4 presents the spectral fluxes of emergent radiation
at a local surface point,
$F_\nu=\int_0^1 \mu I_\nu {\rm d}\mu$ (at $y=0$),
for several effective temperatures and chemical compositions
(pure hydrogen, helium, and iron),
together with blackbody spectra at the same values of $\tef$.
The atmosphere model spectra differ substantially from the blackbody ones,
particularly in high-energy tails of the radiation from the
light-element (hydrogen and helium) atmospheres. The reason is in the 
combination of two effects: rapid decrease of the light-element
opacities with energy, $k_\nu\sim E^{-3}$, and temperature growth
in the surface layers, $T(y)$, with depth $y$.
Hence, the high-energy
radiation is formed in deeper and hotter layers, with $T>\tef$.
The spectra emitted from the heavy-element atmospheres
(see also Zavlin \& Pavlov 2002 for spectra of solar-mixture
compositions) exhibit numerous spectral
lines and photoionization edges (e.g., M, L, and K
spectral complexes
in the iron spectra, at about 0.1, 0.8, and 7.1 keV, respectively)
produced by ions in various ionization stages.
Generally, they are closer to the blackbody
radiation because the energy dependence
of the heavy-element opacities
is, on average, flatter than that for the light elements.

Although the opacity of the
atmospheric plasma is isotropic in the nonmagnetic case,
the emitted radiation show substantial anisotropy,
i.~e., the specific intensity $I_\nu$ depends on the direction
of emission due to the limb-darkening effect
(see the right panel in Fig.~4):
the larger angle $\theta$ between the normal to the surface and 
direction of a specific intensity is, the longer path throughout the
surface layers emerging photons travel to escape, 
and, hence, the stronger absorption the
intensity undergoes.
The anisotropy depends on photon energy and chemical composition
of the atmosphere. This
effect should be taken
into account
to model thermal radiation from a nonuniform
neutron star surface.

The emergent radiation depends also on the surface gravity:
a stronger gravitational acceleration
increases the density of the atmospheric plasma,
changes temperature run $T(y)$
and enhances the nonideality effects,
which results in
weaker (more smoothed)
spectral features. The hardness of the spectral
Wien tail at higher photon energies
also alters with varying surface gravity because
of the changes in the atmosphere structure
(Zavlin, Pavlov \& Shibanov 1996, Heine et al. 2006).
However, these effects
are rather subtle and may be imortant
only for analyzing observational data of extremely good statistics.

\subsection{Magnetized atmosphere models}

First magnetized hydrogen models have been
developed
by Shibanov et al.~(1992), Pavlov et al.~(1994), Shibanov \& Zavlin (1995),
Pavlov et al.~(1995), and Zavlin et al.~(1995)
These models used
simplified radiative opacities of strongly magnetized,
partially ionized plasma, which
did not include the
bound-bound transitions.
However, they
are considered to be
reliable enough in the case of high temperatures,
$\tef\approx 1$ MK, at typical pulsar fields,
$B\sim 10^{12}$ G, when the atmospheric
plasma is almost fully ionized even in the strong
magnetized fields.
Later on, completely ionized hydrogen models for superstrong magnetic fields,
$B\sim 10^{14}$--$10^{15}$ G, have been
presented in a number of papers
(Bezchastnov et al.\ 1996,
 \"Ozel 2001, Ho \& Lai 2001,
Zane et al.\ 2001, Ho \& Lai 2003),
concerned mainly with the vacuum polarization
effects first discussed by Pavlov \& Gnedin (1984)
and the proton cyclotron
lines whose energies shift into the X-ray band at $B\gapr 2\times 10^{13}$ G.
Ho et al. (2003) presented models for partially ionized 
hydrogen atmospheres with magnetic fields up to $5\times 10^{14}$ G
and effective temperatures down to about 0.5 MK.
This work showed that the vacuum polarization affects not only
the proton cyclotron line but also it supresses spectral
features caused by bound species, making them virtually unobservable
in thermal spectra of strongly magnetized neutron stars.
First set of magnetized atmospheres with a heavy-element
composition (pure iron) was constructed by Rajagopal,
Romani \& Miller (1997), with the use of a rather crude
approximations for the very complicated properties of iron ions in
strong magnetic fields. Recently, a next step in modeling
magnetized heavy-element
(carbon, oxygen, neon) atmospheres with $B=10^{12}$--10$^{13}$ G
and $\tef=(1$--5) MK has been undertaken by Mori \& Ho (2006).
These models imply latest developments in atomic physics and
radiative opacities in
strong magnetic fields (Mori \& Hailey 2002, 2006).
Like in the nonmagnetic case, the magnetized heavy-element atmosphere
emission shows many
prominent spectral features which, if observed 
in real X-ray observational data, could be very useful
to measure the neutron star magnetic field and mass-to-radius
ratio, $M/R$.

All the above-mentioned works used
the same approach for  constructing magnetized atmosphere
models, which is generally similar to the nonmagnetic case.
The main difference
is that the atmospheric radiation is polarized, and the
radiative opacities depend on the polarization and direction of
radiation. Gnedin \& Pavlov (1974) described the radiative transfer
in a strongly magnetized plasma in terms of coupled
equations for specific intensities of
two normal modes,
$I_{\nu,1}$ and
$I_{\nu,2}$, with different polarizations and
opacities:

$$
\mu\frac{{\rm d}}{{\rm d}y} I_{\nu,j}(\vec{n}) =
k_{\nu,j}(\vec{n}) I_{\nu,j}(\vec{n}) - 
$$
$$
- \left[\sum_{i=1}^2\oint\, {\rm d}\vec{n'}\,  
I_{\nu,i}(\vec{n'})\, \sigma_{\nu,ij}(\vec{n'},\vec{n})
+ \alpha_{\nu,j}(\vec{n})\frac{B_\nu}{2}\right],~~~~~~(5)
$$ 
where $\vec{n}$ is the (unit) wave-vector,
$\alpha_{\nu,j}$ is the absorption opacity for the $j$-th
mode, $\sigma_{\nu,ij}$ is the scattering opacity
from mode $i$ to mode $j$, 
and $k_{\nu,j}=\alpha_{\nu,j}+
\sum_{i=1}^2\,\oint\,{\rm d}\vec{n'}\,\sigma_{\nu,ij}(\vec{n'},\vec{n})$
is the total opacity. 
It should be noted that the opacity
depends 
on the angle between the wave-vector and
the magnetic fiel,
 so that $I_\nu$ depends not only on $\theta$
but also on $\Theta_{\rm B}$, the angle between
the local magnetic field and the normal to the surface element.
Similar to the nonmagnetic case, Eqs.~(5) are
supplemented with the equations of
hydrostatic and radiative equilibrium (for the latter, Eq.~[4] applies
with $I_\nu=I_{\nu,1}+I_{\rm,2}$).

To deal with the problems
caused by the sharp angular dependence
of the radiative opacities (Kaminker, Pavlov \& Shibanov 1982),
a two-step method for 
modeling of magnetic neutron star
atmospheres was developed (Pavlov et al.~1994;
Shibanov \& Zavlin 1995). At the first step, the radiative transfer
is solved in the diffusion approximation for the mean
intensities $J_{\nu,j}=(4\pi)^{-1}\, \oint I_{\nu,j}(\vec{n})\, {\rm d}\vec{n}$: 
\setcounter{equation}{5}
\begin{equation}
\frac{{\rm d}}{{\rm d}y} d_{\nu,j} \frac{{\rm d}}{{\rm d}y} J_{\nu,j}=
\bar{\alpha}_{\nu,j}\left[J_{\nu,j}-\frac{B_\nu}{2}\right]+
\bar{\sigma}_\nu\left[J_{\nu,j}-J_{\nu,3-j}\right]\, ,
\end{equation}
where
$\bar{\alpha}_{\nu,j}=(4\pi)^{-1}\oint\,{\rm d}\vec{n}\,  
\alpha_{\nu,j}(\vec{n})$ and 
$\bar{\sigma}_\nu=(4\pi)^{-1}\oint\!\oint\,{\rm d}\vec{n}\,{\rm d}\vec{n'}\,
\sigma_{\nu,12}(\vec{n},\vec{n'})$ are the angle-averaged absorption
and scattering opacities. The diffusion coefficient is
$d_{\nu,j}=d_{\nu,j}^{\rm p}\,\cos^2\Theta_{\rm B}+
d_{\nu,j}^{\rm o}\,\sin^2\Theta_{\rm B}$, with
$d_{\nu,j}^{\rm p}=\int_0^1\,\mu^2\,k_{\nu,j}^{-1}\,{\rm d}\mu$ and
$d_{\nu,j}^{\rm o}=\int_0^1\,(1-\mu^2)\,k_{\nu,j}^{-1}\,{\rm d}\mu$.
Next, the atmospheric structure obtained at the first step
is corrected using an iterative procedure applied to the
exact equations of the radiative transfer.
Finally, the emergent intensity (at $y=0$) is 
\begin{equation}
I_{\nu,j}=\mu^{-1}\,\int_0^\infty\,\left[\alpha_{\nu,j}\,
\frac{B_\nu}{2}+\sum_{i=1}^2\,\sigma_{\nu,ij} J_{\nu,i}\right]\,
\exp\left[-\nu^{-1}\,\int_0^y\,k_{\mu,j}\,{\rm d}z\right]\,{\rm d}y\,,
\end{equation}
and the emitted spectral flux is computed as 
$F_\nu=\int_0^1\,\mu\,\sum_{i=1}^2\,I_{\nu,i}\,{\rm d}\mu$. More
details on the modeling of magnetized atmospheres can be found in
Pavlov et al. (1995).

\begin{figure}
\centerline{\psfig{file=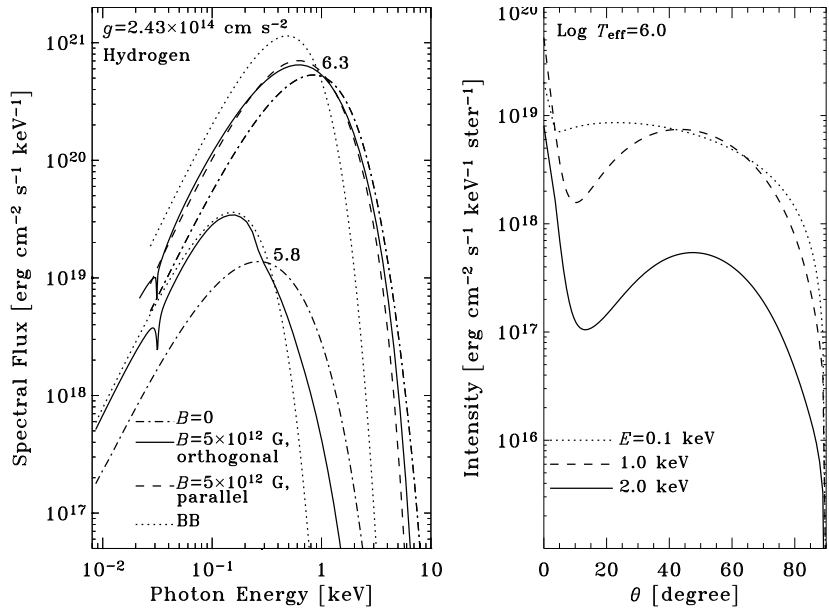,height=8cm,clip=} }
\caption{
{\sl Left}: Spectra of radiation emergent
from neutron star atmospheres for a magnetic
field orthogonal and parallel to the surface
with different effective temperatures (numbers near
the curves label $\Log\tef$ [$\tef$ in K]).
`BB' stands for blackbody spectrum.
{\sl Right}: Dependences of specific intensities on
the angle between the photon wave-vector and the magnetic
field directed along the surface normal.
}
\label{fig:5}
\end{figure}

Figure~5 (left panel) shows polarization-summed spectral
fluxes of the emergent radiation, 
$F_\nu$,
emitted by a local element of the neutron star surface,
for two values of effective temperature and two 
magnetic field orientations,
perpendicular and parallel to the surface
($\Theta_{\rm B}=0$ and 1, respectively).
The main result is that the magnetized atmosphere spectra
are harder than the blackbody radiation of
the same $\tef$, although they are softer than the nonmagnetic
spectra.
Similar to the nonmagnetic
case, this is explained
by the temperature growth with depth and
the opacity decrease
at higher energies, which is more gradual
($\propto E^{-1}$ for the mode
with smaller opacity) in the magnetized plasam.
At lower effective temperatures, $\tef\lapr 1$ MK,
the photoionization opacity (due to bound-free transitions)
becomes important, that affects the shape of emitted
spectra (see the example with $\Log\tef=5.8$ in Fig.~5).
The proton cyclotron lines are seen at energies
$E=6.3\,(B/10^{12}\,{\rm G})$ eV.
If the magnetic field is very large, $B\gapr 10^{14}$ G,
the proton cyclotron line shifts into the X-ray band.
On the other hand, if 
the magnetic field is not so large, $B=10^{10}$--$10^{12}$ G,
the neutron star atmosphere spectra may exhibit
the electron cyclotron lines
in the X-ray band,
at $E_{\rm ce}=11.6\, (B/10^{12}\, {\rm G})$ keV.
Calculations of hydrogen atmosphere
models which include bound-bound transitions (Zavlin \& Pavlov 2002,
Ho et al. 2003)
show that spectral lines, considerably broadened by
the motional Stark effect (Pavlov \& M\'esz\'aros 1993, Pavlov
\& Potekhin 1995)
may emerge at $\tef\lapr 0.5$ MK. The strongest line
corresponds to the transition between the ground state and
the lowest excited state; its energy is
$E\approx [75+0.13 \ln(B/10^{13}\, {\rm G}) + 
63(B/10^{13}\, {\rm G})]$ eV.

Radiation emerging
from a magnetized  atmosphere is
strongly anisotropic. Angular dependences of
the local specific intensities, $I_\nu=I_{\nu,1}+I_{\nu,2}$ (Eq.~[7]),
show a complicated ``pencil-plus-fan'' structure ---
a narrow peak along the direction of the magnetic field
(where the atmospheric plasma is most transparent for
the radiation),
and a broader peak at intermediate angles. The widths
and strengths of the peaks depend on magnetic field and photon energy 
(see examples in the right panel of Fig.~5).
Obviously, it is very important
to account for this anisotropy while modeling the
radiation from 
a neutron star with nonuniform surface
magnetic field and effective temperature.

\subsection{Thermal radiation as detected
by a distant observer}

Results presented in \S\S\,3.3 and 3.4 describe spectral 
radiation emitted by a {\sl local} element at the neutron star surface.
The effective temperature and/or magnetic 
field distributions over the surface can be nonuniform
(for example, if a neutron star has a dipole magnetic field,
the effective temperature 
decreases from the magnetic poles to the equator). 
To calculate
the {\sl total} emission, one has to integrate
the local intensities, computed for 
local temperatures
and magnetic fields, over the visible part of the
surface $S$, with account for the 
gravitational
redshift and bending of photon trajectories:

\begin{equation}
F(E_{\rm obs}) = \gr\frac{1}{D^2}\int_S \mu\, 
I(E_{\rm obs}/\gr)\, {\rm d}S\, , 
\end{equation}
where  $E_{\rm obs}=\gr\,E$ is the observed (redshifted) 
photon energy.
To take into account the interstellar absorption,
a factor, $\exp[-n_{\rm H} \sigma_{\rm eff}(E)]$,
should be added in Eq.~(8)
($\sigma_{\rm eff}[E]$ is the absorption cross section per
hydrogen atom).
More details about the integration over the 
neutron star surface  can be found in Pavlov \& Zavlin (2000b).
It is worthwhile to mention
that if a neutron star has a nonuniform distribution
of the magnetic field, the integration
broadens the spectral features. In addition,
if a neutron star is a fast rotator, one should take into account
the Doppler shifts of energies of
photons emitted from surface elements moving with
different radial velocities.
Maximum values of these velocities, 
$v_r=2\pi R P^{-1} \sin\zeta$ 
($\zeta$ is the inclination of the rotation axis with respect to 
observer's line of sight --- see Fig.~3), 
can be as high as 10\%--15\% of the speed 
of light for millisecond periods. For instance,
Zavlin \& Pavlov (2002) showed that a fast
rotation, $P\lapr 10$ ms, may lead to complete smearing of 
weak and narrow spectral lines, provided $\sin\zeta$ is large enough,
leaving only most prominent spectral jumps around the strongest
photoionization edges.

If thermal radiation originated from small polar caps on the
neutron star surface, it greatly simplifies Eq.~(8):
\begin{equation}
F(E)=g_{\rm r}\frac{S_{\rm a}}{D^2}
I(E/g_{\rm r},\, \theta^*)\, ,
\end{equation}
where the apparent spot area $S_{\rm a}$ and the angle
$\theta^*$ between the wave-vector of escaping radiation 
and the radius-vector to the hot spot are computed
with account for the effect of gravitational 
bending.
These quantities depend on
the angles $\alpha$
(between the rotational and magnetic axes)
 and $\zeta$ (Fig.~3), and
the gravitational parameter $\gr$
(see Zavlin, Shibanov \& Pavlov
1995 for details).

The flux given by Eqs.~(7) or (8)
varies with the period of neutron star rotation.
One can obtain a large variety of 
pulse profiles 
at different assumptions on the 
angles $\alpha$ and $\zeta$ 
and 
the neutron star mass-to-radius ratio.
Examples of pulse profiles computed for radiation
from the entire neutron star surface are shown
in Zavlin \& Pavlov (2002), 
whereas pulse profiles of thermal radiation from heated polar
caps are presented by Zavlin, Shibanov \& Pavlov (1995)
and Zavlin \& Pavlov (2004a) for magnetized atmosphere models,
and by Zavlin \& Pavlov (1998) for nonmagnetic ones.

\subsection{Atmosphere emission vs. blackbody radiation}
Although the model atmosphere spectra are different
from the blackbody radiation,
very often an observed thermal spectrum
can be fitted equally well with the blackbody and neutron star atmosphere
models (see examples in \S\,4), 
particularly when the energy resolution is low and/or
the energy band is narrow and/or observational data are of a poor
quality.
However, the parameters obtained from such fits are quite different, 
especially when the hydrogen or helium atmospheres are used.
Since the light-element atmosphere spectra are much harder
than the blackbody spectra
at the same effective temperature,
atmosphere model fits
result in temperatures $T_{\rm atm}$ significantly lower
than the blackbody temperature $T_{\rm bb}$,
with a typical ratio $T_{\rm bb}/T_{\rm atm}\sim 2$--3.
On the other hand, to provide the same total energy flux,
the blackbody fit yields a smaller 
normalization factor, proportional to $S/D^2$ (see Eq.~[7]),
than the  atmosphere model fit does.
In other words, the light-element atmosphere
fit gives a considerably larger size
of the emitting region, $S_{\rm atm}/S_{\rm bb}\sim 50$--200,
for the same distance to the source. Note however that
both neutron star atmosphere and blackbody models yield
about the same values of bolometric luminosity $\lbol^\infty=\gr^2\lbol$
as measured by a distant observer.

It is also worth to remember that blackbody radiation is 
{\sl isotropic} and, hence, it results in
weak pulsations of model flux, with a typical pulsed fraction 
around a few percents only.

Finally, the atmosphere models discussed here, both nonmagnetic
and magnetized, are available for analyzing thermal
X-ray emission observed from neutron stars
as a part of the X-ray Spectral
Fitting Package\footnote{
{\tt http://heasarc.gsfc.nasa.gov/docs/xanadu/xspec/}} 
(codes `NSA' and `NSAGRAV' in XSPEC)
provided by the NASA's High Energy Astrophysics Science
Archive Reserach Center.

\subsection{Modeling radiation from condensed neutron star surface}
As mentioned above, if magnetic field of the neutron star surface is 
strong enough and the surface temperature is rather low,
then the outermost surface layers could be
in any state other than gaseous.
For example, at $B=10^{14}$ G and 
$\ts\lapr 0.5$ MK and $\ts\lapr 2$ MK for hydrogen and iron compositions,
respectively, the surface would be in a condensed (solid) state 
(van Adelsberg et al. 2005). First models of thermal radiation
emitted by condensed surface of a netron star were constructed by 
Turolla, Zane \& Drake (2004) and van Adelsberg et al. (2005).
These works showed that the overall spectral shape of X-ray flux
emitted by a condensed surface is mostly featureless (only weak
spectral features associated with ion cyclotron
and electron plasma frequencies can appear in some cases) and
fairly close to blackbody spectrum of the same temperature.
The main difference between these two model spectra is
that, because of suppressed emissivity of condensed surface, 
the surface radiation is reduced from the blackbody
one by a factor of a few. Hence, applying condensed surface models to
observed thermal emission is expected to result in temperature 
estimates close to and flux normalizations
(proportional to the  factor $[R^\infty/D]^2$) larger by a factor of few 
than those yielded by blackbody radiation.

\section{Thermal emission from neutron stars: observational
results}

As already mentioned in \S\,2, thermal emission has been observed
from a rather large number of neutron stars of different types.
The majority of them is radio pulsars of different ages ranging
from very young neutron stars to old and very old (millisecond) ones.
In addition to active pulsars, a number of radio-quiet
neutron stars emitting only thermal-like X-rays have been detected,
with typical temperatures $\sim 0.5$--5 MK.
They are usually subdivided in four classes: Anomalous X-ray
Pulsars (AXPs; Mereghetti et al. 2002, Kaspi 2006),
Soft Gamma-ray Repeaters (SGRs; Kaspi 2004),
``dim'' or ``truly isolated'' radio-quiet neutron starss 
(i.e., not associated with
SNRs; Haberl 2007) and compact
central sources (CCOs) in SNRs (Pavlov et al. 2002, Pavlov,
Sanwal \& Teter 2004) which have been
identified with neither active pulsars nor AXPs/SGRs.
Observational manifestations
(particularly, multiwavelength spectra) of
radio-quiet neutron stars are quite different from those of
active pulsars, and their properties 
have not been investigated as extensively,  
but the presence of the thermal
component in their radiation provides a clue to understand the nature
of these objects. While the paper by Pavlov, Zavlin \& Sanawal
(2002) provides a detailed review on thermal
emission from neutron stars, here I discuss a few most
interesting and illustrative examples, concentrating mainly on 
spectral properties of detected thermal emission.

\begin{figure}
\centerline{\psfig{file=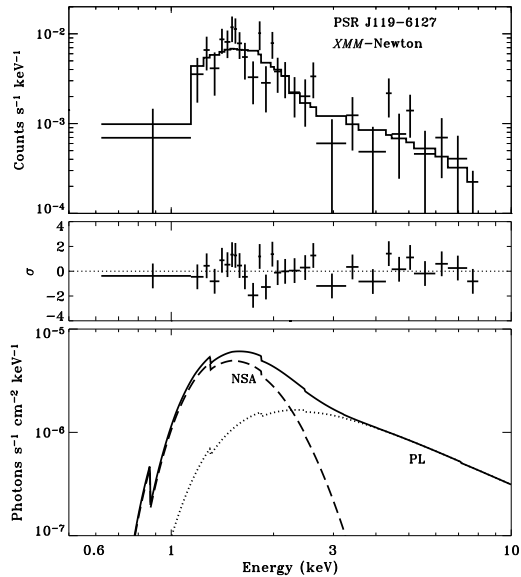,height=8cm,clip=} }
\caption{Two-component, hydrogen magnetized atmosphere (NSA)
model plus power law (PL), fit to the X-ray spectrum
of PSR J1119--6127 detected with \xmm\ (upper panel). 
The middle panel shows
residuals in the fit, whereas the lower panel presents
the contributions 
(attenuated by interstellar absorption)
from the thermal (dashes) and nonthermal (dots) components
(see \S\,4.1).
}
\label{fig:6}
\end{figure}

\subsection{PSR~J1119-6127}
It is the youngest\footnote{
The characteristic age, $\tc=P/(2\dot{P})$, is
a standard age estimate for the vast majority of radio and X-ray
pulsars. Note however that it may be very inaccurate
(see \S\,4.3).},
$\tc=1.6$ kyr,  and very energetic in terms of the spin-down power,
$\ed=2.3\times 10^{36}$ ergs s$^{-1}$, radio pulsar whose 
X-ray flux reveals a strong thermal component.
The best representation of the pulsar's spectrum
detected with \xmm\ is a two-component, thermal plus nonthermal, model
(Gonzalez et al. 2005).
The nonthermal emission dominating at energies $E\gapr 2.5$ keV
is well fitted with a power-law spectrum of a photon index
$\Gamma\simeq 1.5$ and X-ray luminosity 
$\lnon\simeq 0.8\times 10^{33}$ ergs s$^{-1}$ in the 0.2--10 keV 
range\footnote{
This energy range is used for all other estimates on $\lnon$
given in this work.}.
The thermal component can be fitted with a blackbody spectrum of
an apparent (redshifted, see \S\,3.1) temperature 
$\tbb\simeq 2.4$ MK and radius $\rbb\simeq 3.4$ km (for the estimated
distance\footnote{Distances
cited in \S\,4 are either those estimated to SNRs
which host some of discussed objects or
derived from pulsar parallaxes  or 
dispersion measures.}
to the pulsar $D=8.4$ kpc), implying the measured bolometric
luminosity $\lbol^\infty\simeq 2.7\times 10^{33}$ ergs s$^{-1}$.
This model fit would mean that the thermal radiation originates
from a small hot area on the pulsar's surface (polar caps?),
although the inferred radius of the emitting area significantly
exceeds the canonical radius $\rpc^*\simeq 0.2$ km predicted by theoretical
models for PSR~J1119--6127 with a
spin period $P=0.408$ s (see \S\,2).
Alternatively, the thermal component can be interpreted as
X-ray flux of an effective (actual)
temperature $\tef\simeq 1.6$ MK (or $\tefin=\gr^{-1}\tef\simeq 1.2$ MK)
emitted from the whole pulsar's surface covered with a magnetized
($B\approx 1\times 10^{13}$ G) hydrogen atmosphere\footnote{Note
that the parameters of the atmosphere model cited in this work
differ from those given in Gonzalez et al. (2005).} (assuming the 
standard neutron star
mass $M=1.4\,M_\odot$ and radius $R=10$ km), yielding 
$\lbol\simeq 4.7\times 10^{33}$ ergs s$^{-1}$ (or 
$\lbol^\infty=\gr^2\lbol\simeq 2.8\times 10^{33}$ ergs s$^{-1}$, very close
to the value obtained in the blackbody fit).
In this interpretation, the parameters of the nonthermal component
are virtually the same as those in the fit with the blackbody radiation.
The best fit with the magnetized atmosphere model and power-law
spectrum is shown in Figure~6.

Very importantly, the X-ray flux of PSR~J1119--6127 detected in
the 0.6--2 keV range, where the thermal component dominates, is
pulsed, with a very large pulsed fraction, $\fp\approx 75$\%
(Conzalez et al. 2005). It should be noted that
because of the  strong gravitational bending effect (\S\,3.2)
such pulsations can be 
reconciled with neither (isotropic) blackbody 
radiation nor atmospheric
emission from an {\sl uniform} surface. On the other hand, no pulsations
have been detected at energies $E>2$ keV, that is rather a surprising
result as nonthermal emission is expected to be strongly pulsed,
especially that emitted by young and energetic pulsars.

The example of PSR~J1119-6127 is remarkable in the sense
that the situation with observing thermal emission from
very young and active pulsars is not in fact as ``pessimistic'' 
as it may follow from the general picture described in \S\,2,
and more such detections can be expected in future.

\begin{figure}
\centerline{\psfig{file=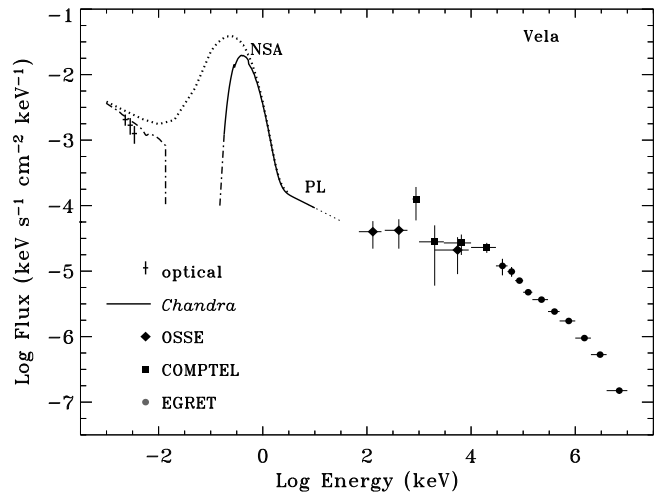,height=8cm,clip=} }
\caption{Multiwavelength spectrum of the Vela pulsar
detected with different missions.
The solid line shows the X-ray spectrum 
obtained with \chan\ and fitted with a two-component, 
neutron star atmosphere (NSA) and power-law (PL), model (see \S\,4.2).
Dots correspond to the unabsorbed model spectrum. 
The dash-dotted lines show the extrapolated
optical and EUV absorbed spectra.
}
\label{fig:7}
\end{figure}

\subsection{The Vela pulsar and PSR~B1706-44}
The superb angular resolution of \chan\ made it possible to 
separate X-ray flux of the famous Vela pulsar ($P=0.089$ s,
$\tc=11$ kyr, $\ed=6.9\times 10^{36}$ ergs s$^{-1}$)
from its bright pulsar-wind nebula and study the properties
of the pulsar's emission (Pavlov et al. 2001).
The \chan\ observations revealed that the bulk of the X-ray
flux detected from Vela is of a thermal origin, and nonthermal
emission dominates only at energies $E\gapr 2$ keV, similar
to the case of PSR~J1119--6127. The thermal component
can be described equally well with either a blackbody spectrum
or a magnetized
($B\approx 5\times 10^{12}$ G)
hydrogen\footnote{The featureless spectrum of Vela
obtained with \chan\ at a high-energy resolution indicate 
that there are no heavy elements 
on the pulsar's surface.}
atmosphere model. However, the parameters of the thermal
component are significantly different in the blackbody 
and atmosphere model fits: $\tbb\simeq 1.6$ MK and 
$\rbb\simeq 2.8$ km,
$\tef\simeq 0.9$ MK (or $\tefin\simeq 0.7$ MK) and $R\simeq 13$ km
(for the estimated distance to the pulsar $D=300$ pc), respectively.
The bolometric luminosity of the thermal emission is
$\lbol\simeq 0.8\times 10^{33}$ ergs s$^{-1}$.
Moreover, the slope of the nonthermal emission depends 
on the thermal model applied for interpreting the pulsar's spectrum.
It is a rather large photon index $\Gamma\simeq 2.7$ 
if the blackbody radiation is used.
The nonthermal component with this slope greatly exceeds the optical
emission of the pulsar.
In the analysis involving the atmosphere model the nonthermal component
is much flatter, with $\Gamma\simeq 1.5$.
Remarkably, the extrapolation of this power-law spectrum 
(with $\lnon\simeq 0.2\times 10^{32}$ ergs s$^{-1}$,
or about 40 times lower than $\lbol$)
matches fairly well the optical and hard X-ray/soft $\gamma$-ray
fluxes detected from the pulsar. This is shown in Figure~7.

The X-ray pulsed profile of Vela is very unusual and complicated,
with at least three peaks per rotational period and $\fp\approx 8$\%(Pavlov, 
Zavlin \& Sanwal 2002). A combined spectral and timing analysis
is crucial to further elucidate mechanisms generated
X-ray emission of this pulsar. 

PSR~1706--44 is one more young and energetic pulsar ($P=0.102$ s, $\tc=18$ kyr,
$\ed=3.4\times 10^{36}$ ergs s$^{-1}$) emitting thermal X-rays, 
with spectral properties very similar 
to those of Vela (McGowan et al. 2004). 
The thermal component of  PSR~1706--44 detected with \xmm\
can be described by a magnetized hydrogen atmosphere model
with $\tef\simeq 1.0$ MK and $R\simeq 12$ km (for $D=2.3$ kpc),
or $\lbol\simeq 1.0\times 10^{33}$ ergs s$^{-1}$.
The nonthermal emission is fitted with a power-law
spectrum of $\Gamma\simeq 1.4$, but its luminosity, 
$\lnon\simeq\lbol$, is much higher than that of Vela.
The X-ray pulsed profile of PSR~1706--44 is energy-dependent
and shows a broad pulse per period with $\fp\approx 10$\%
at energies $E\lapr 1.4$ keV,
where the thermal flux dominates.

\begin{figure}
\centerline{\psfig{file=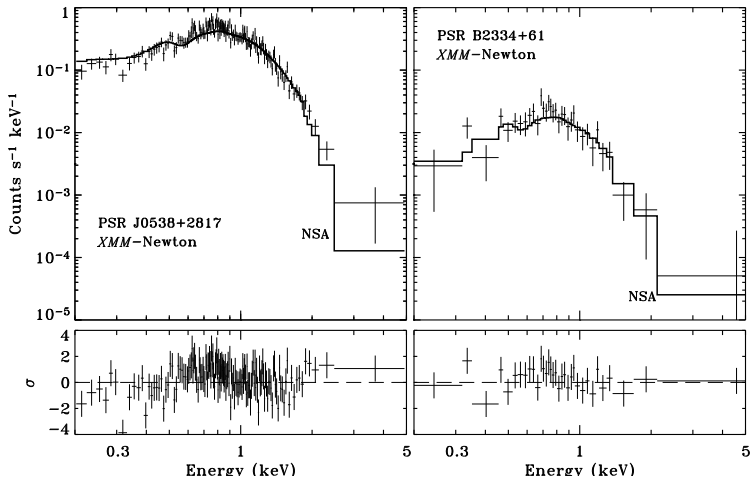,width=10cm,clip=} }
\caption{X-ray spectrum of PSRs J0538+2817 and B2334+61 detected
with \xmm\ (crosses) and fitted
with magnetized neutron star atmosphere models (see \S\,4.3). Residuals in
the fits are shown in the lower panel.
}
\label{fig:8}
\end{figure}

\subsection{PSRs J0538+2817 and B2334+61}
These two neutron stars have similar pulsar parameters (age, spin period,
spin-down power) and could be considered as ``twins'', or 
at least ``coevals'', if their ages were derived in the same way.

PSR~J0538+2817 ($P=0.143$ s, $\ed=4.9\times 10^{34}$ ergs s$^{-1}$) 
represents a rare case when neutron star age
is well determined. Based on the pulsar's
proper motion measurements, Kramer et al. (2003) inferred
the true age of PSR~J0538+2817, $\tau\simeq 30$ kyr, much smaller than 
the standard estimate $\tc=618$ kyr. 

No nonthermal emission was observed from PSR~J0538+2817.
The pulsar's spectrum detected with \xmm\ can be fitted
with a single blackbody radiation of $\tbb\simeq 2.1$ MK and
$\rbb\simeq 1.7$ km  for 
a distance $D=1.2$ kpc (McGowan et al. 2003). 
As shown by Zavlin \& Pavlov (2004b),
a hydrogen atmosphere model with
$B=10^{12}$ G fits the observational data even better,
yielding the surface temperature $\tef\simeq 1.1$ MK and
the pulsar radius $R\simeq 10.5$ km (at $M=1.4 M_\odot$),
or $\lbol\simeq 1.2\times 10^{33}$ ergs s$^{-1}$.
An upper limit on luminosity of a possible nonthermal
component is $\lnon<1.0\times 10^{31}$ ergs s$^{-1}$
(assuming $\Gamma=1.5$).

The X-ray flux of PSR~J0538+2817 is pulsed, with a broad, 
asymmetric pulse per period and 
pulsed fraction $\fp\approx 25$\%.
The phases of pulse maxima at energies below and above 0.8 keV
differ by $\sim 75^\circ$ (Zavlin \& Pavlov 2004b). This
indicates that the thermal emission is intrinsically anisotropic and
the pulsar has a strong nonuniformity of
the surface temperature and magnetic field.

The estimate on the age of PSR~B2334+61 ($P=0.495$ s, 
$\ed=6.2\times 10^{34}$ ergs s$^{-1}$) is obtained in the
standard way, $\tc=41$ kyr. Similar to the case of PSR~J0538+2817,
the X-ray flux of PSR~B2334+61 detected with with \xmm\
if of a thermal origin, and the pulsar's spectrum can be
fitted with a single thermal model (McGowan et al. 2006).
The blackbody fit yields $\tbb\simeq 1.5$ MK and
$\rbb\simeq 2.8$ km for 
a distance $D=3.1$ kpc. 
A hydrogen atmosphere models with $R=10$ km, $M=1.4 M_\odot$ and 
$B=10^{13}$ G fits equally well the observational data,
resulting in the surface temperature $\tef\simeq 0.9$ MK
and $\lbol\simeq 0.5\times 10^{33}$ ergs s$^{-1}$.
A lower limit on luminosity of a possible nonthermal
component is $\lnon<0.7\times 10^{31}$ ergs s$^{-1}$ (for $\Gamma=1.5$).
Based on the results of the spectral fits, one can assume that these
two pulsars are indeed ``twins'' and the estimate $\tc$ on the age
of PSR~B2334+61 is close to the pulsar's true age. Figure~8 shows the 
spectra detected from PSRs J0538+2817 and B2334+61 and 
fitted with the best neutron star atmosphere models. The only difference
in the X-ray properties of these two objects is that the emission 
observed from  PSR~B2334+61 revealed no pulsations, with a 5\%
upper limit on the pulsed fraction, indicating different 
neutron star geometries of these pulsars (e.g., 
PSR~B2334+61 could have smaller $\zeta$ and/or $\alpha$ angles --- see Fig.~3).

\subsection{Middle-aged pulsars: B0656+14, B1055--52, and Geminga}
As discussed in \S\,2, middle-aged (a few hundred thousand 
years old) pulsar are believed to be best targets for observing thermal
neutron star emission. The well-known three neutron stars
with close pulsar parameters, PSRs
B0656+14, B1055--52, and Geminga\footnote{
Dubbed as ``Three Musketeers'' by Joachim Tr\"umper.}, 
support this. Observations with 
$ROSAT$ first showed that soft X-ray emission from these objects
are of a thermal origin (\"Ogelman 1995), and later \chan\ and \xmm\
allowed a detailed study of this radiation (Pavlov, Zavlin \& Sanwal 2002,
Zavlin \& Pavlov 2004b, De Luca et al. 2005, Kargaltsev et al. 2005).

\begin{figure}
\centerline{\psfig{file=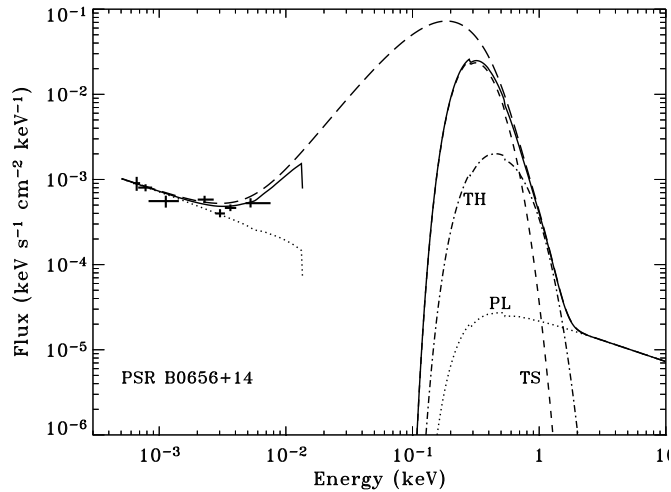,height=8cm,clip=} }
\caption{Broadband spectrum of PSR~B0656+14 for a three-component model
(TS+TH+PL; see \S\,4.4) extrapolated in optical.
The solid and long-dashed curves show the absorbed and unabsorbed
spectra, respectively. Crosses indicate the IR-optical fluxes.
}
\label{fig:9}
\end{figure}

PSR~B0656+14 ($P=0.385$ s, $\tc=111$ kyr, $\ed=3.8\times 10^{34}$ ergs s$^{-1}$)
is the brightest of these three neutron stars. Its X-ray spectrum cannot be
fitted by a two-component model like those describing the spectra
of PSRs J1119--6127, B1706--44,
and Vela. If fitted with a blackbody radiation,
the pulsar's thermal emission requires two components,
a ``soft'' one (TS) with  $\tbbs\simeq 0.8$ MK and $\rbbs\simeq 7.5$ km,
and a ``hard'' one (TH) with $\tbbs\simeq 1.7$ MK and $\rbbs\simeq 0.6$ km
(for $D=300$ pc). The TS component of the
bolometric luminosity $\lbols^\infty\simeq 1.6\times 10^{32}$ ergs s$^{-1}$
may be regarded as emitted from the
whole pulsar's surface, whereas the TH component 
($\lbols^\infty\simeq 0.2\times 10^{32}$ ergs s$^{-1}$)
could be interpreted as radiation from heated polar caps.
In addition to these two thermal components, a power-law spectrum is
needed to fit the pulsar's emission detected at energies above 2 keV. 
With the available data, the slope of the nonthermal component is
not well constrained, but one can assume
that the photon index does not change from optical to X-rays,
like in the Vela pulsar. Then, it
results in a power-law spectrum with $\Gamma\simeq 1.5$ and 
$\lnon\simeq 0.3\times 10^{31}$ ergs s$^{-1}$. Figure~9 presents
the broadband emission of PSR~B0656+14.

The X-ray spectrum of PSR~B1055--52 ($P=0.197$ s, $\tc=535$ kyr,
$\ed=3.0\times 10^{34}$ ergs s$^{-1}$) is very similar to that
of PSR~B0656+14. It can be fitted only with a three-component
model, ``soft'' and ``hard'' blackbody radiation plus a power law,
with the following parameters (as inferred by Pavlov,
Zavlin \& Sanwal 2002 from the combined $ROSAT$ and \chan\ data on the pulsar):
$\tbbs\simeq 0.8$ MK and $\rbbs\simeq 8.4$ km,
$\tbbs\simeq 1.6$ MK and $\rbbs\simeq 0.6$ km (for $D=700$ pc),
a photon index $\Gamma\simeq 1.7$ and 
$\lnon\simeq 0.9\times 10^{31}$ ergs s$^{-1}$.

\begin{figure}
\centerline{\psfig{file=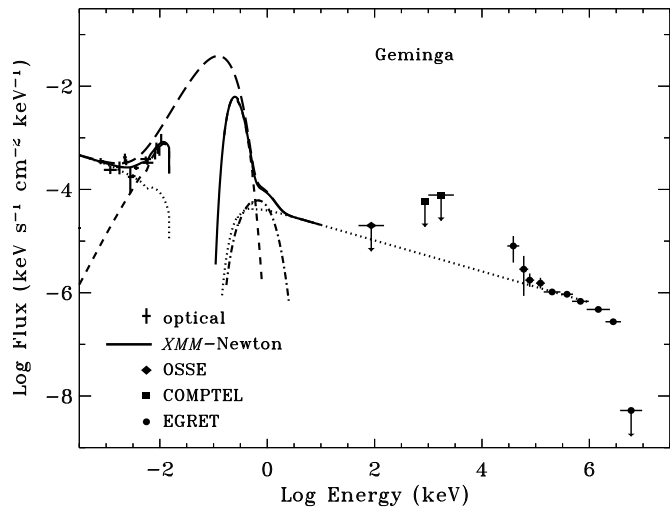,height=8cm,clip=} }
\caption{Multiwavelength spectrum of Geminga observed with
different missions. The X-ray flux
is described with a three-component (TS+TH+PL) model.
The solid and long-dashed curves are the absorbed and 
unabsorbed fluxes, respectively. Short dashes, dash-dots, and
dots show the TS, TH, and PL components, respectively
(see \S\,4.4 ).
}
\label{fig:10}
\end{figure}

Compared to the spectra of PSRs B0656+1 and B1055--52,
the X-ray flux of the famous $\gamma$-ray and X-ray
Geminga pulsar\footnote{
Geminga has not been firmly confirmed as a radio pulsar.}
($P=0.237$ s, $\tc=342$ kyr,
$\ed=3.3\times 10^{34}$ ergs s$^{-1}$) alone does not require
a three-component model. It can be fitted with two components,
a blackbody spectrum and a power law of $\Gamma\simeq 2.0$. However,
the nonthermal component extrapolated to low energies greatly
exceed optical fluxes observed from the pulsar.
To describe the optical/$UV/FUV$ and X-ray data with the same model, 
one needs to invoke a three-component interpretation of the X-ray flux,
similar to that applied for the other ``Musketeers'', with
$\tbbs\simeq 0.5$ MK and $\rbbs\simeq 12.9$ km,
$\tbbs\simeq 2.3$ MK and $\rbbs\simeq 0.05$ km (for $D=200$ pc),
and a photon index $\Gamma\simeq 1.5$ and 
$\lnon\simeq 0.2\times 10^{31}$ ergs s$^{-1}$
(Kargaltsev et al. 2005).
Figure~10 shows the multiwavelength spectrum of Geminga based on this
three-component interpretation. 
It is worthwhile to mention that,
although this spectral model is similar to those suggested
for the X-ray emission of PSRs B0656+1 and B1055--52,
the $\rbbh$ radius inferred
for Geminga is smaller by about a factor of 10 than the estimates
obtained for the other two pulsars. Note that according to theoretical pulsar
models (\S\,2) these three objects should have about the same
polar cap radii.

Applying magnetized hydrogen atmosphere models
for the thermal components observed from all three pulsars
yields formally acceptable fits. However, they imply
very large radii for the TS component, $R\gapr 40$ km.
Therefore, applicability of the available neutron star
atmosphere models to these objects is questionable.

The pulsations of the X-ray fluxes from these
three pulsars
shows a complex behavior, with energy-dependent
variations in pulsed fraction, phase of main pulses,
and pulse shape (Pavlov, Zavlin \& Sanwal 2002,
Zavlin \& Pavlov 2004b, Kargaltsev et al. 2005).
This indicates that their thermal radiation is
locally anisotropic, in obvious contradiction
with the simplistic blackbody interpretation of the phase-integrated
spectra. Moreover, the observed pulsed profiles
hints that the surface distributions of
temperature and magnetic field are not azimuthally symmetric,
suggesting a strong multipolar component of the magnetic field
or a decentered magnetic dipole. 

\subsection{Old radio pulsars}
Because of their age, $\tc>1$ Myr, old ordinary 
(with spin periods $P\gapr 0.05$ s, i.e., not millisecond) 
radio pulsars are expected to be
and actually are much less energetic and fainter than their 
younger ``stellarmates''. Up to now, of about 1,100 
such pulsars known\footnote{According the pulsar catalog
provided by the Australia Telescope National Facility;
{\tt http://www.atnf.csiro.au/research/pulsar}\,.},
only seven have been firmly detected in X-rays\footnote{
Marginal X-ray detection of two more old pulsars have been reported
(Zavlin \& Pavlov 2004a).}
(Zavlin \& Pavlov 2004a,
Zhang, Sanwal \& Pavlov 2005, Kargaltsev, Pavlov \& Garmire 2006).
The analysis of X-rays collected from these old neutron stars
revealed very diverse properties of their emission, with thermal
radiation undoubtedly detected from two objects, PSRs B0950+08 and
J2043+2740 (Zavlin \& Pavlov 2004a).

\begin{figure}
\centerline{\psfig{file=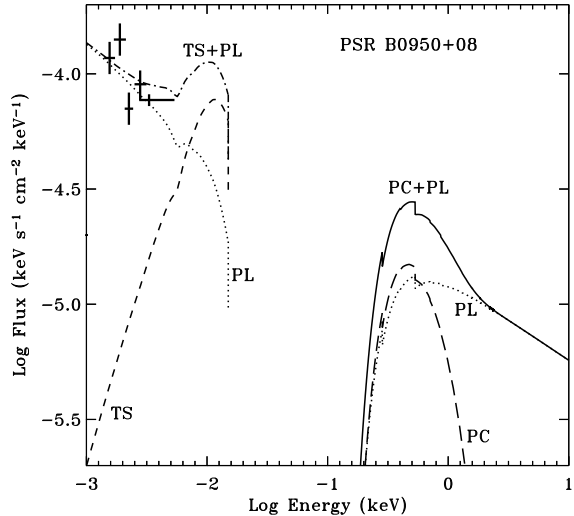,height=7cm,clip=} }
\caption{Broadband spectrum of PSR~B0950+08 for a two-component,
polar caps (PC) plus power law (PL), model
(see \S\,4.5) extrapolated in optical.  
Crosses show the optical fluxes. Radiation from
the whole surface (TS) is also indicated.
}
\label{fig:9}
\end{figure}

\subsubsection{PSR~B0950+08}
The X-ray spectrum of PSR~B0950+08 ($P=0.253$ s, $\tc=17.4$ Myr, 
$\ed=5.6\times 10^{32}$ ergs s$^{-1}$) detected with \xmm\
is best described with a two-component model, thermal plus
nonthermal. The thermal component, dominating
at energies $E\lapr 0.7$ keV, is interpreted as radiation
from two heated polar caps on the star's surface covered with 
a magnetized ($B\approx 3\times 10^{11}$ G) hydrogen atmosphere.
The applied model takes into account the GR effects (redshift
and gravitational bending). 
The inferred temperature, radius, and luminosity of the polar caps
are $\tpc\simeq 1.0$ MK, $\rpc\simeq 0.25$ km,  
and $\lbolpc\simeq 0.3\times 10^{30}$ ergs s$^{-1}$
(for a distance $D=260$ pc), respectively.
Remarkably, the obtained polar cap radius is in excellent
agreement with the conventional estimate $\rpc^*\simeq 0.3$ km
(\S\,2).
The nonthermal emission is fitted with a power-law spectrum of
a photon index $\Gamma\simeq 1.3$ and luminosity 
$\lnon\simeq 1.0\times 10^{30}$ ergs s$^{-1}$.
This power-law model also matches well optical fluxes
detected from the pulsar. Figure~11 presents the broadband, from
optical to X-rays, spectrum of PSR~B0950+08. 
The analysis of the temporal behavior of the pulsar's X-ray
flux, with energy-dependent pulse shape and pulsed fraction,
also supports this two-component interpretation.
The combined optical and X-ray data put the upper limit
on the temperature of the bulk of the neutron star
surface, $\ts<0.1$ MK (assuming the standard neutron star radius $R=10$ km).

\subsubsection{PSR~J2043+2740}
Analysis of the X-ray flux of PSR~J2043+2740 ($P=0.096$ s, $\tc=1.2$ Myr, 
$\ed=5.6\times 10^{34}$ ergs s$^{-1}$) observed with \xmm\ firmly showed,
despite a low number of photons collected,
that the pulsar's spectrum is very soft, with no emission detected
at energies $E\gapr 2$ keV. A single power-law fit to these data yields
a photon index $\Gamma\simeq 4.7$, that greatly exceeds a typical value
$\Gamma=1$--2 found in nonthermal radiation of a large sample of
radio pulsars (including the examples discussed in this paper),
with ages varying in a broad range, from about 1 kyr to 20 Myr.
This fact completely rules out a nonthermal interpretation of
the X-ray emission of PSR~J2043+2740. Applying blackbody radiation to
these X-ray data yields
$\tbb\simeq 0.9$ and $\rbb\simeq 2.7$ km (for a distance $D=1.8$ kpc),
that could be suggestive that the X-ray emission originates
from polar caps. However, this radius estimate is a factor of
5 larger than the theoretical prediction $\rpc^*\simeq 0.5$ km.
On the other hand, the fits with magnetized 
($B\approx 4\times 10^{11}$ G) hydrogen atmosphere models gives
the surface effective temperature $\tef\simeq 0.6$ MK
for the neutron star radius $R=9$ km. The latter fit indicates
that the detected X-ray emission most likely emerges from the
bulk of the star's surface, with the bolometric luminosity 
$\lbol\simeq 0.8\times 10^{30}$ ergs s$^{-1}$. This result
is rather unexpected because PSR~J2043+2740 has the highest
spin-down power among all known ordinary pulsars with 
$\tc>1$ Myr and, hence, it should have been the strongest
nonthermal emitter among old ordinary pulsars. 
Hopefully, someday a longer observation
of this pulsar will provide more details on the properties
of its thermal X-ray emission. 

\subsection{Millisecond pulsars}
Millisecond pulsars, with unique properties, represent an
evolutionarily distinct group among radio pulsars. First of
all, they possess very short and stable spin periods, $P\lapr 0.05$ s
with $\dot{P}\lapr 10^{-18}$ s s$^{-1}$, and low surface magnetic fields, 
$B\lapr 10^{10}$ G.
They are thought to be extremely old neutron stars 
($\tc\sim 0.1$--10 Gyr) presumably spun up by angular momentum
transfer in binary systems. 
X-ray detections have been reported for
about 35 (nonaccreting) millisecond pulsars
(of more than a hundred currently known). 
The majority of them are located in the globular cluster
47~Tuc and exhibit thermal X-rays most probably
emitted from heated polar caps (Bogdanov et al. 2006).
However, detailed spectral and timing information on X-ray emission
has been obtained only for eight of the detected millisecond pulsars
(see Zavlin 2007 for a review). One half of them are nonthermally emitting
pulsars. The bulk of X-rays from the other four objects
originates from heated polar caps. These are PSRs J0030+0451,
J2124--3358, J1024--0719, and J0437--4715, with similar
characteristics of the detected X-ray flux. The latter is the 
nearest ($D=140$ pc) and brightest millisecond pulsar, and
properties of its X-ray emission are discussed below.

\begin{figure}
\centerline{\psfig{file=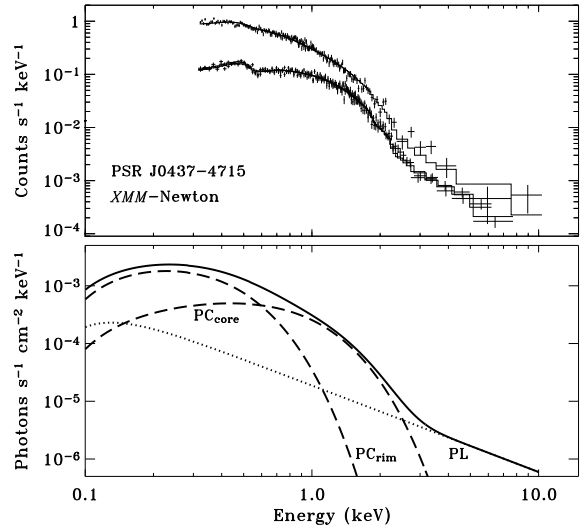,height=7cm,clip=} }
\caption{X-ray spectrum of PSR~J0437--4715 detected with
different instruments onboard \xmm\ (crosses in the upper panel)
and fitted with a composite model, two-temperature (``core'' and ``rim'') 
polar caps (PC) and power law (PL), shown in the lower panel 
(see \S\,4.6).
}
\label{fig:12}
\end{figure}

\subsubsection{PSR J0437--4715}
Pulsed X-ray emission from this pulsar 
($P=5.76$ ms, $\tc=6.5$ Gyr, $\ed=3.8\times 10^{33}$ ergs s$^{-1}$)
was discovered with $ROSAT$
(Becker \& Tr\"umper 1993), and observations with \chan\ and 
\xmm\ have finally established its properties (Zavlin et al. 2002,
Zavlin 2006). The model describing the pulsar's X-ray flux consists
of a thermal and nothermal components.
The thermal component is emitted from two identical
polar caps covered with 
a (nonmagnetic) hydrogen atmosphere and located at the poles
of a magnetic dipole. 
As first proposed by Zavlin \& Pavlov (1998), the polar caps 
of a millisecond pulsar would have
a nonuniform temperature because low surface magnetic field does not 
prevent the energy (heat) released by relativistic particles from
propagating along the surface to an area of a radius larger than
the conventional estimate $\rpc^*$. The uniform temperature
is approximated by a step-function mimicking a smaller and hotter
polar cap ``core'' and a larger and colder ``rim''. The GR effects 
(redshift and gravitational bending) are accounted for
in this interpretation. The thermal model, supplemented with 
a power-law component, fits well the X-ray emission detected from 
PSR~J0437--4715 up to 10 keV and yields reasonable spectral parameters:
$\tpcc\simeq 1.4$ MK and $\tpcr\simeq 0.5$ MK,
$\rpcc\simeq 0.4$ km and $\rpcr\simeq 2.6$ km, with the total
bolometric luminosity $\lbolpc\simeq 1.8\times 10^{30}$ ergs s$^{-1}$.
The nonthermal component has a photon index  $\Gamma\simeq 1.8$
and luminosity $\lnon\simeq 0.5\times  10^{30}$ ergs s$^{-1}$.
Figure~12 presents this model and the fit to the data on
PSR~J0437--4715 collected with \xmm. Interestingly,
PSR~J0437--4715 has beeen
detected in $UV/FUV$ with $HST$ (Kargaltsev, Pavlov \& Romani 2004).
The shape of the inferred spectrum suggests thermal emission
from the whole neutron star surface of a surprisingly high
temperature of about 0.1 MK. A powerful energy source
(most likely, internal chemical and/or frictional heating) should be
operating in a Gyr-old neutron star to keep its surface
at such temperature.

X-ray emission from all four thermally emitting millisecond pulsars 
is pulsed, with pulsed fraction $\fp\simeq 35$\%--50\% (Zavlin 2007).
Such pulsed fraction can be produced only by intrinsically
anisotropic radiation, that supports the assumption on presence
of a hydrogen atmosphere on the surface of millisecond pulsars. 
The pulsed profiles of PSRs J0437--4715, J2124--3358, and
J1024--0719 are rather similar in shape, with single
broad pulses,
whereas the light curve of PSR~J0030+0451
exhibits two pulses per period indicating that the geometry of
this pulsar 
(the angles $\zeta$ and $\alpha$ --- see Fig.~3)
is different from those of the three others.
For example, in the framework of the conventional pulsar model
with the magnetic dipole at the neutron star center,
PSR~J0030+0451 can be a nearly orthogonal rotator
(i.e., $\zeta\simeq\alpha\simeq 90^\circ$)
with two pulses in its light curve being due to contributions
from two polar caps seen during the pulsar's rotation. For the others,
the bulk of the detected X-ray flux is expected to
come mostly from one polar cap. Importantly, as first demonstrated
by Pavlov \& Zavlin (1997) and Zavlin \& Pavlov (1998) on the 
X-ray emission of PSR~J0437--4715 detected with $ROSAT$,
analyzing pulsed emission with thermal polar cap models can put
stringent constraints
on the neutron star mass-to-radius ratio $M/R$ if the star's
geometry is known (e.g., from radio polarization data).

\begin{figure}
\centerline{\psfig{file=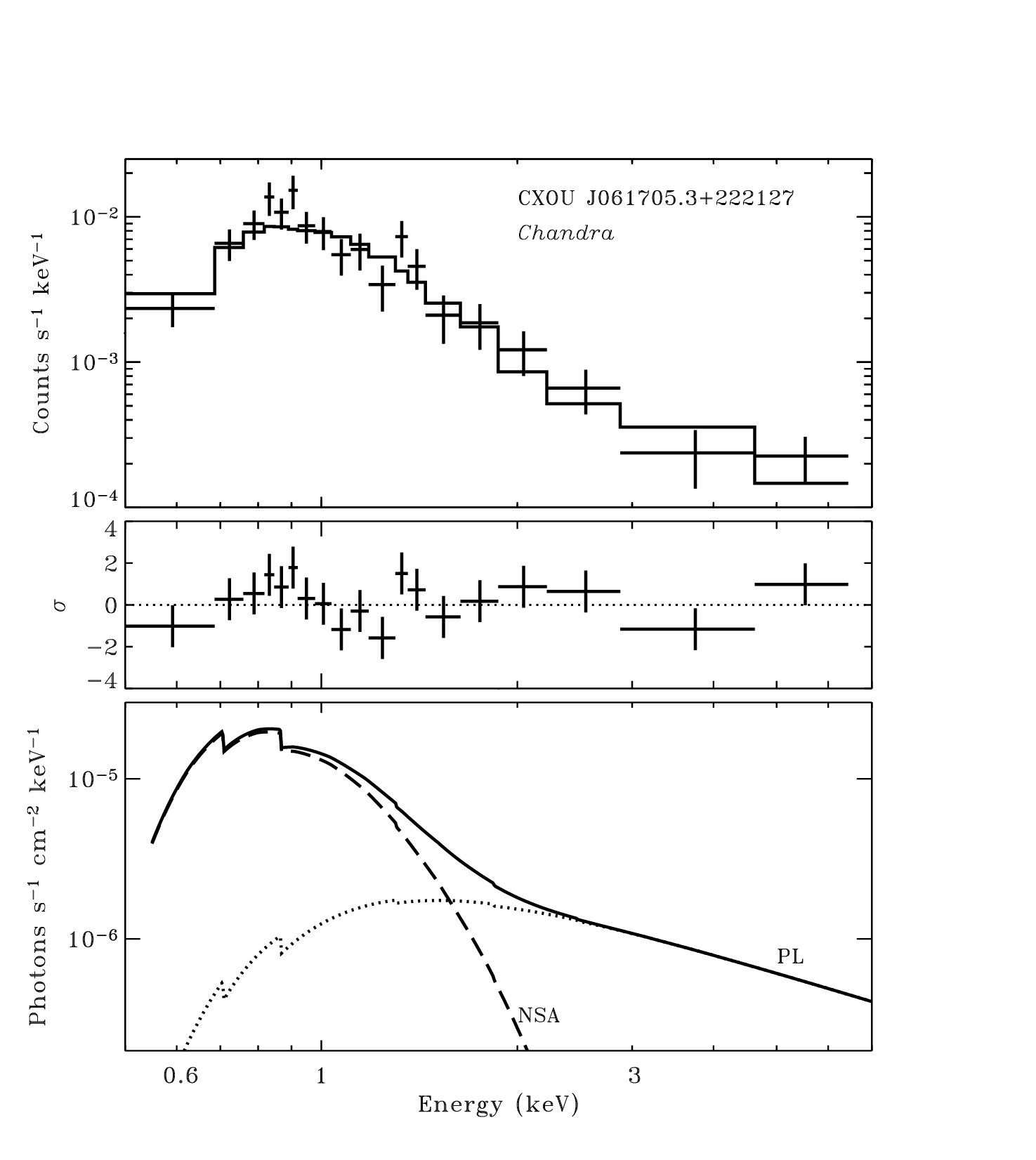,height=8cm,clip=} }
\caption{Two-component, hydrogen magnetized atmosphere (NSA)
model plus a power law (PL), fit to the X-ray spectrum
of CXOU~J061705.3+222127 detected with \chan\ (upper panel). 
The middle panel shows
residuals in the fit, whereas the lower panel presents
the contributions 
(attenuated by interstellar absorption)
from the thermal (dashes) and nonthermal (dots) components
(see \S\,4.7).
}
\label{fig:13}
\end{figure}

\subsection{Putative pulsars: CXOU~J061705.3+222127 ($=$ J0617)
and RX~J0007.0+7302 (=J0007)}
The compact source J0617 discovered in a short \chan\ 
observation (Olbet et al. 2001)
is located within a bright X-ray comet-like nebula. 
Most likely, J0617 is a young, fast and energetic pulsar
that powers this nebula. To firmly confirm this
very plausible hypothesis, pulsations 
of emission from this object 
(in radio and/or X-rays) have to be detected yet.
A longer \chan\ observation of J0617 and the nebula
provided more details on X-ray properties of the source
and surrounding diffuse emission
(Gaensler et al. 2006, Weisskopf et al. 2007).
The X-ray spectrum of J0617
reveals a thermal component which dominates at energies
$E\lapr 1.7$ keV. At higher energies, a nonthermal
emission prevails. The fact that the spectrum of J0617
is very similar to those found in the young and powerful pulsars,
J1119--6127 (\S\,4.1), Vela, and B1706--4 (\S\,4.2), 
strongly supports the assumption
on this compact source being a neutron star and a pulsar.
The detected spectrum can be equally well fitted with both blackbody
plus power-law and hydrogen atmosphere plus power-law combinations.
Applying magnetized atmosphere models interprets the thermal flux
as emitted from the whole neutron surface of 
$\tef\simeq 0.8$ MK and radius $R=10$ km (for a  
distance $D=1.5$ kpc), with $\lbol\simeq 2.9\times 10^{32}$ ergs s$^{-1}$,
and yields 
the nonthermal spectrum of $\Gamma\simeq 1.2$ with
$\lnon\simeq 0.2\times 10^{32}$ ergs s$^{-1}$,  
about 15 times smaller than the thermal luminosity.
The spectrum of J0617 fitted with this two-component model is presented
in Figure~13.
It should be noted that, like in the case of
Vela, using blackbody radiation instead of
atmosphere models results in much steeper power-law component
of $\Gamma\simeq 2.7$, that is not
typical for nonthermal emission from radio pulsars. Hence, the
interpretation involving the atmosphere models can be regarded as
more preferable.

Another putative pulsar with a possible $\gamma$-ray counterpart
powering an X-ray nebula
is the compact source J0007 at the center of the SNR CTA~1
(but not a CCO discussed in \S\,4.8). 
As obtained by Slane et al. (2004b),
its X-ray spectrum detected with \xmm\ is well fitted
with a magnetized hydrogen atmosphere model 
of the same parameters as those derived for J0617 (assuming
$D=1.4$ kpc), plus
a power-law component of $\Gamma\simeq 1.6$ and 
$\lnon\simeq 0.5\times 10^{32}$ ergs s$^{-1}$. 
Extrapolation of this power-law spectrum to high
energies is consistent with the flux detected
from the proposed $\gamma$-ray counterpart,
strengthening the proposition that J0007 is a $\gamma$-ray emitting
pulsar.

\subsection{``Pure'' thermally emitting neutron stars}
All objects presented in \S\S\,4.1--4.7 are either
radio pulsars or show other manifestations
of the nonthermal activity. Below I briefly discuss a few examples 
of radio-quiet neutron stars emitting only thermal X-ray emission. 

\begin{figure}
\centerline{\psfig{file=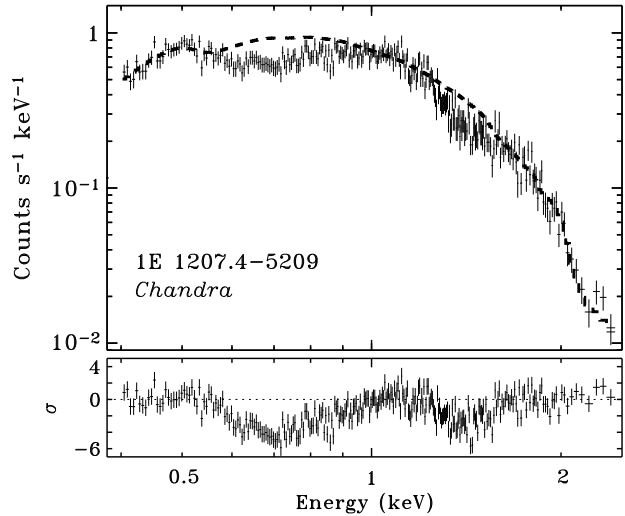,height=7cm,clip=} }
\caption{{\it Upper panel}: Spectrum of 1E 1207.4--5209 
detected with \chan\ (crosses) vs. a featureless thermal
model (dashes). 
{\it Lower panel}: Residuals between the observed and model
spectra demonstrating 
the presence of the two absorption features in
the X-ray emission of this object (see \S\,4.8).
}
\label{fig:14}
\end{figure}

\subsubsection{1E~1207.4--5209 ($=$ 1E1207) and other CCOs}
1E1207 belongs to the small group of 
currently known seven CCOs in SNRs (see Pavlov, Sanwal \& Teter 2004 for 
a review). One of them, the CCO in the SNR RCW~103,
is very outstanding because it shows a highly variable X-ray flux and its 
emission is presumably powered by accretion
from a companion in a close binary system with a $\sim 6.5$-hr
orbital period. The other six CCOs have not
shown any long-term variability of their thermal emission,
characterized by blackbody temperatures $\tbb\approx 2$--5 MK and 
emitting areas $\rbb\approx 0.3$--3 km, and seem to be
similar to each other. However, the spin periods\footnote{
Typical for radio pulsars.}
of two objects, 
1E1207 in the SNR PKS~1209--51/52 with $P=0.424$ s (Zavlin et al. 2000) and 
CXOU~J185238.6+004020 in the SNR Kes~79  with $P=0.105$ s (Gotthelf,
Halpern \& Seward 2005), make them distinct from the rest.

1E1207 is even more unique: it is the only known nonaccreting neutron star
whose X-ray flux contains two firmly detected spectral features.
Figure~14 presents the spectrum of 1E1207 with two absorption lines
at about 0.7 and 1.4 keV discovered with \chan\
(Sanwal et al. 2002). These features cannot be explained as proton or 
electron cyclotron lines, and  
current interpretations 
involve atomic transitions of once-ionized helium in a very
strong magnetic field, $B\sim 10^{14}$ G (Sanwal et al. 2002; Pavlov \&
Bezchastnov 2005),
or transitions of helium-like oxygen (or neon) 
in a weaker field $B\sim 10^{11}$ G (Mori \& Hailey 2002).
First magnetized oxygen atmosphere models of Mori \& Ho (20006)
seem to be in apparent qualitative agreement with the X-ray spectrum
of 1E1207, although it has to be demonstrated yet whether these models
could explain the observational data in the quantitative
way. In any case, regardless of what the true
origin of these spectral lines is, they 
make 1E1207 one of the most important objects
for astrophysics of neutron stars and physics of superdense matter
because it provides an opportunity to measure the gravitational 
redshift at the neutron star surface and
constrain the equation of state of the superdense matter in the 
neutron star interiors.

\subsubsection{``Truly isolated'' X-ray emitting neutron stars}
The final part of the observational section gives
a short description of another small group of very
intriguing objects --- so-called ``dim''\footnote{
As discussed at the {\sl ``Isolated Neutron Stars: 
from the Interior to the Surface''} conference held in April 2006 in
London (UK), the historical name ``dim'' is rather inappropriate because
some of these objects are very bright X-ray emitters.}
or ``truly isolated'' neutron stars. These objects,
seven in total, were discovered with $ROSAT$, and a lot of important
information on properties of their X-ray emission provided with
\chan\ and \xmm\ is summurized in the detailed reviews by
Haberl (2007) and van Kerkwijk \& Kaplan (2007).
All seven emit thermal X-ray spectra characterized by
temperatures $\tbb\approx 0.7$--1.1 MK. Faint optical counterparts 
(with magnitudes $m_B>25$) were
identified for five of them. Four objects have periods in the 3.5--11.4 s range,
and spin period candidates in the same range have been proposed
for the other three. All this strongly suggests that these objects are 
neutron stars (and I believe nobody doubts this).
Extrapolations of the observed X-ray spectra fitted with a blackbody
spectrum to optical strongly underpredict measured optical fluxes
(where available; see Pavlov, Zavlin \& Sanwal 2002
for examples). None of current neutron star atmosphere models
applied to the X-ray spectra of these objects
either fits or yields reasonable parameters (generally, the atmosphere
models result in unrealistic estimates on the neutron stars size, that
in turn leads to a large overestimate of observed optical fluxes ---
see Pavlov et al. 1996).
Timing solutions (spin period derivatives) determined for two objects
yield very similar estimates on the stars' age, $\tc\approx 2$ Myr,
spin-down power, $\ed\approx 4\times 10^{30}$ ergs s$^{-2}$,
and surface magnetic field, $B\approx 3\times 10^{13}$ G.
The low estimates derived on $\ed$ can explain
the absence of nonthermal activity of these two (and, by analogy, all other
five) neutron stars. 
Broad absorption features centered at energies 
in the 0.3--0.8 keV range have been
detected in the radiation of six objects, whereas the seventh one reveals
almost a ``perfect'' blackbody  spectrum (Pavlov, Zavlin \& Sanwal 2002).
What produces the absorption features in the X-ray spectra
of six objects is a matter of debate:
they might be caused by proton cyclotron resonance in a magnetic field
$B\gapr 5\times 10^{13}$ G or produced by atomic transitions in, 
for example, a strongly magnetized hydrogen surface (if it is in the gaseous
state). Any of these hypotheses
should be taken with caution until reliable models of surface emission
are proposed for these neutron stars. 
In this respect, the model by P\'erez-Azorin et al.
(2006) invoking emission from a nonuniform, strongly
magnetized and condensed neutron star surface seems to be the most
advanced and self-consistent approach to explain
emission properties of thse objects.
There are other intriguing
details on properties of these objects (e.g., a long-term variability
of the spectral shape of X-ray flux detected from one of them,
presumably cased by a neutron star precession) that can be found
in the above-mentioned reviews.

\section{Concluding remarks}
I would like to complete this review with a brief discussion on
what has been learned during the extensive, 15-year-long
studying of thermal emission from isolated neutron stars.

Undoubtedly, a substantial progress has been made on the theoretical front.
Best investigated models are nonmagnetic atmospheres of
various chemical compositions and magnetized fully-ionized hydrogen
atmospheres. These models have been successfully applied to 
interpretation of thermal emission to a number of neutron stars,
mainly, radio pulsars of different ages, including millisecond pulsars,
and yielded reasonable neutron star parameters (surface temperatures
and radii of emitting areas).
Besides the active pulsars, there is a group of 
neutron stars transiently accreting in X-ray binaries
(e.g., Aquila~X-1, KS~1731--260, Centaurus~X--4, 4U~1608--522,
MXB~1659--29, 4U~2129+47)
whose X-ray emission in quiescence has been
analyzed with use of the atmosphere models (Rutledge et al. 1999, 2001a,b
and 2002, Nowak, Heinz \& Begelman 2002,
Wijnands et al. 2002 and 2003,
Heinke et al. 2006). Although these objects are not isolated,
their quiescent radiation is interpreted as emitted from the whole
neutron stars surface covered with a nonmagnetic hydrogen atmosphere
heated by energy released in pycnonuclear reactions of the compressed
accreted material. Importantly, based on the results obtained on the
thermal emission from these objects, Yakovlev, Levenfish \& Haensel (2003)
proposed a new method for studying neutron stars internal structure
and equation of state of the inner matter. In addition, as suggested
by Rutledge et al. (2000), the atmosphere models can be useful for 
distinguishing between transiently accreting neutron stars and black holes,
in quiescence.
First steps have been undertaken in modeling partially ionized 
atmosphere models with strong magnetic fields and different chemical 
compositions,
as well as in modeling thermal emission from condensed neutron star
surfaces, although both these types of models are still awaiting 
application to observational data.

Despite a lot of interesting and encouraging results
obtained in the thermal emission modeling and with applying these
models to observational data,
a number of problems remains to be solved. First of all, 
the approach based on two
polarization modes currently used in magnetized atmosphere models
is in fact inaccurate and inapplicable for a partially ionized plasma.
To construct more advanced models, the problem of radiative transfer
in strongly magnetized plasmas should be solved in terms of the four Stokes
parameters, with use of the polarizability tensor constructed with 
aid of the Kramers-Kronig relations (Bulik \& Pavlov 1995, Potekhin et al.
2004). Next, 
investigations of the structure of various atoms, molecules,
and molecular chains in strong magnetic fields,
as well as radiative transitions in these species (Pavlov 1998),
are necessary to construct, in combination with the advanced radiative
transfer approach, 
magnetized atmosphere models of a next generation for different
chemical compositions.
Very interesting and important are the
(virtually unknown) radiative properties of matter in
superstrong magnetic fields, $B\gapr 10^{14}$ G,
apparently found in AXPs and SGRs. More reliable models
are required for radiative properties of nonideal plasmas
and condensed matter, as well as further investigations
of phase transitions between different
states of matter in strong magnetic fields.

Not only the models of neutron star atmospheres and condensed surfaces
require improvements. Analysis of observational data on
thermal flux from neutron stars, especially
temporal behavior of detected X-ray emission, shows that 
the idealized picture of a neutron star with a centered magnetic
dipole and uniform surface temperature is oversimplification.
Therefore, future computations of thermal emission from a neutron star
applied to observational data should use realistic surface
temperature distributions to reproduce {\sl both spectral
and temporal} properties of observed emission.
In particular,  the problem of temperature distribution over heated
polar caps of millisecond pulsars is of a special importance because
modeling pulsed thermal emission from these objects is a promising
way to constrain neutron star mass-to radius ratio. For that,
more elaborated models
of magnetospheric pulsed emission are required to disentangle
nonthermal and thermal components.

Confronting the surface temperatures derived from observation data
with theoretical models of neutron star
thermal evolution (Yakovlev et al. 2005) indicates that
the neutron star
interiors are most probably superfluid and that these objects 
may have different masses (e.g., $M\simeq 1.47 M_\odot$ for Vela and Geminga,
and $M\simeq 1.35 M_\odot$ for PSR~B1055--52). But these results are
are quite uncertain because, first of all,
they are based on the assumption that the characteristic age
of a neutron star is its true age (see the example
of PSR~J0538+2817 in \S\,4.3). Next, more importantly, thermal emission 
mechanisms operating in neutron stars
are not completely understood yet, especially
in colder objects. Only rather simple conclusions could be 
drawn from the obtained results.
It looks plausible that younger and
hotter objects (this $\tau\simeq\tc\lapr 50$ kyr and
$\ts\gapr 1$ MK) are  indeed covered with a gaseous atmosphere,
strongly ionized if comprised of hydrogen,
as suggested by the examples discussed in \S\S\,4.1--4.3
and (probably) 4.7. To explain why the simple blackbody model fits well
thermal radiation from colder neutron stars with strong magnetic fields, 
whereas atmosphere models do not provide reasonable parameters (\S\,4.4),
one may suggest that cooling hydrogen-depleted neutron star
envelopes undergo a phase transition, forming a condensed surface. 
But this assumption is challenged by very complicated temporal
behavior of the thermal flux detected from many these objects --- 
it can be hardly
explained without invoking a strong anisotropy of surface radiation
similar to that characteristic to the atmospheric radiation. Therefore,
the parameters inferred from  the blackbody
spectral fits should be taken with caution.
It also concerns the two-blackbody (``soft'' and ``hard'') model
suggested for the thermal phase-integrated spectra of the middle-aged pulsars
(\S\,4.4). It is not clear whether the harder thermal component is real
or it emerges because the simplified spectral models were used
(e.g., this component is not required in the interpretation
involving a power-law spectrum with a phase-dependent photon index ---
see Jackson \& Halpern 2006).

There are even much more unanswered questions related to thermal
radiation of neutron stars (concerning, for example,
the nature of CCOs and connection between them and orther types of
neutron stars, the origin of spectral lines in thermal emission
of seven objects and why no features are present in spectra of
other neutron stars with similar temperatures and magnetic field, etc.)
To answer these questions, not only improved models are necessary but also
a larger sample of neutron stars of various types observed
in different energy ranges, from optical/UV to X-rays, is required. 
In particular, as shown by Kargaltsev et al. (2005)
and Kargaltsev \& Pavlov (2007), the $UV/FUV$ range is very important
for elucidating properties of thermal emission emerging
from the whole neutron star surface. Contrary to the X-ray (Wien)
part of the thermal emission whose shape is strongly affected
by surface chemical composition and temperature inhomogeneties,
the $UV/FUV$ (Rayleigh-Jeans) tail,
proportional to the product $[\ts R^2]$,  
can put tight constraints on the surface temperature.
Hopefully (``the hope dies last'') enough observational
time will be allocated in future for studying these enigmatic
objects, neutron stars.

{\bf\sl Acknowledgement.}
The author gratefully acknowledges the collaboration with George Pavlov
and many other colleagues in studying neutron stars during the last 15
years and thanks the Heraeus Stiftung and the Physikzentrum Bad Honnef 
for support and hospitality.
The author's work at the NASA MSFC is supported
by a NASA Associateship Award.

%

\end{document}